\documentclass[twocolumn,english,aps,superscriptaddress,prb,floatfix,longbibliography]{revtex4-2}
\usepackage[latin9]{inputenc}
\usepackage{color}
\usepackage{babel}
\usepackage{amsmath}
\usepackage{amssymb}
\usepackage{graphicx}
\usepackage{esint}
\usepackage{bm}
\usepackage[unicode=true,pdfusetitle,
 bookmarks=true,bookmarksnumbered=false,bookmarksopen=false,
 breaklinks=true,pdfborder={0 0 0},pdfborderstyle={},backref=false,colorlinks=true]
 {hyperref}
\hypersetup{
 linkcolor=blue,citecolor=blue,urlcolor=blue}

\makeatletter
\usepackage{color}
\usepackage{babel}

\usepackage{color}
\usepackage{dsfont}   %Used to get the identity "1"
\usepackage{braket}
\usepackage[normalem]{ulem}

\newcommand{\nf}{n_{\rm{free}} \left(T\right)}

\makeatother

\begin{document}

\title{Kondo screening and random-singlet formation in highly disordered systems }

\author{Lucas G.  Rabelo}
\affiliation{Instituto de F\'isica, Universidade de S\~ao Paulo, 05315-970, S\~ao Paulo, SP, Brazil}
\author{Igor C. Almeida}
\affiliation{Instituto de F\'isica de S\~ao Carlos, Universidade de S\~ao Paulo, S\~ao Carlos, SP, 13560-970, Brazil}
\author{Eduardo Miranda}
\affiliation{Gleb Wataghin Physics Institute, The University of Campinas, Rua S\'ergio Buarque de Holanda, 777, CEP 13083-859 Campinas, SP, Brazil}
\author{Vladimir Dobrosavljevi\'c}
\affiliation{Department of Physics and National High Magnetic Field Laboratory, Florida State University, Tallahassee, FL 32306}
\author{Eric C. Andrade}
\affiliation{Instituto de F\'isica, Universidade de S\~ao Paulo, 05315-970, S\~ao Paulo, SP, Brazil}

% Vlad: I will make my comment and inserts in blue

\begin{abstract}

We introduce a minimal effective framework designed to address the anomalous low-temperature thermodynamics observed in strongly disordered interacting systems near a metal-insulator transition, such as doped semiconductors. Our approach examines how a strongly inhomogeneous electronic bath quenches dilute pairs of local magnetic moments. By solving the resulting two-impurity Kondo-Heisenberg problem within a large-$\mathcal{N}$ variational framework, we capture the competition between Kondo screening and random-singlet formation. Both mechanisms give rise to singular thermodynamic responses, producing a magnetic susceptibility of the form $\chi(T)\sim T^{-\alpha}$, where the exponent $\alpha$ evolves smoothly with model parameters and saturates in the metallic regime, consistent with experimental observations. Our results establish a unified framework that connects random-singlet physics and disorder-induced Kondo screening in strongly disordered correlated materials.

\end{abstract}
\date{\today}
\maketitle

\section{Introduction} 

Doped semiconductors with a shallow impurity band, for example, Si:P,  provide a unique setup for studying the metal-insulator transition (MIT) \citep{lee_ramakrishnan,shklovskii,paalanen91,nfl_review05,lohneysen11,vlad12},  which is driven by varying the dopant concentration.  Because the dopant ions are strongly positionally disordered, one might expect the MIT to exhibit Anderson-like behavior; however, the experimentally observed critical exponents are dramatically inconsistent with the non-interacting model.  Furthermore, both the magnetic susceptibility and the specific heat anomalies indicate the presence of local magnetic moments on the insulating side, as well as across the entire transition region \citep{paalanen91,lakner89}. This hints that Coulomb repulsion is key to localizing the impurity electrons, strongly suggesting a Mott-like scenario, which is decisively modified by disorder \citep{paalanen88, roy88, milovanovic89, bhatt92, byczuk05, carol09, amd09}.  Therefore, the MIT in doped semiconductors combines Anderson and Mott localization, posing a truly challenging theoretical problem \citep{lee_ramakrishnan,vlad12,abanin19}.

In the insulating regime, the low-energy magnetic properties are well captured by the Bhatt-Lee random-singlet model \cite{bhatt_lee82}. Here, disorder generates a wide range of antiferromagnetic exchange couplings, leading to a hierarchical condensation of spin pairs into singlets as temperature decreases. This process results in characteristic power-law thermodynamic singularities \citep{paalanen88,paalanen91,nfl_review05,lohneysen11}. Conversely, on the metallic side, local moments are predominantly screened by itinerant electrons. Strong disorder makes the conduction environment highly inhomogeneous, creating broad distributions of local Kondo temperatures and again yielding singular thermodynamic responses \cite{vlad92a,miranda01,nfl_review05,cornaglia06,kaul07,kettemann09,kettemann12,miranda14,andrade15}. 

These two mechanisms --- random-singlet formation and disorder-induced Kondo screening --- have often been discussed separately.  However,  close to a metal-insulator transition, they are intertwined, making a unified understanding both essential and challenging. In this work, we present a minimal effective model that bridges this gap. Drawing on the phenomenological two-fluid picture of doped semiconductors \cite{paalanen88, paalanen91}, which postulates both localized and delocalized electrons near the MIT because of the very broad disorder distribution, we analyze the low-temperature thermodynamics of diluted pairs of local magnetic moments in strongly disordered systems.

\begin{figure}[t]
\begin{centering}
\includegraphics[width=1\columnwidth]{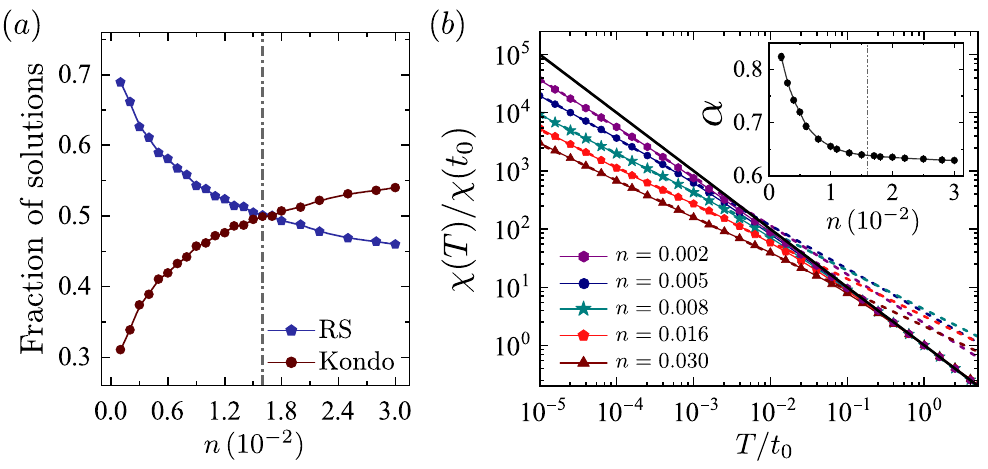}
\par\end{centering}
\caption{\label{fig:kondo2} Two-site Kondo-Heisenberg problems embedded in a disordered bath with varying impurity concentrations $n$.  (a)  Fraction of solutions as a function of $n$.  The pairs either form a random singlet (RS) or are Kondo screened.   (b) Magnetic susceptibility $\chi\left(T\right)$ as a function of $T$,  on a log-log scale,  for several $n$.  The full black line is the Curie law.  The dashed lines are power-law fits: $\chi\left(T\right)\propto T^{-\alpha}$.  (Inset) Exponent $\alpha$ as a function of $n$. The vertical dot-dashed line marks the MIT for the conduction bath at $n_c=0.016$.  We considered $L=30$.}
\end{figure}

Specifically, we study an ensemble of two-impurity Kondo-Heisenberg systems using a large-$\mathcal{N}$ variational method. We find that a fraction of local-moment pairs form random singlets, while the rest are Kondo-screened, generating an inhomogeneous local Fermi-liquid phase.  The balance between these populations shifts from a singlet-dominated state in the insulator to a Kondo-dominated state in the metal [Fig.~\ref{fig:kondo2}(a)].  Strikingly,  the model predicts a singular low-temperature susceptibility, $\chi(T) \sim C/T \sim T^{-\alpha}$, with the exponent $\alpha$ varying smoothly with dopant concentration $n$ [Fig.~\ref{fig:kondo2}(b)], consistent with experiments on heavily doped semiconductors ~\citep{andres81, paalanen91,nfl_review05,lohneysen11,vlad12}.  The resulting ground state is thus highly spatially inhomogeneous, with local moments quenched at widely varying energy scales.

We organize the manuscript as follows.  In Sec.~\ref{sec:model}, we present the effective low-energy model we consider,  discuss its non-interacting limit, and present the variational approach to solve the model in the interacting case.  In Sec.~\ref{sec:RS}, we discuss the effective model deep into the insulating regime,  where we recover the random-singlet phase.  Sec.~\ref{sec:1kondo} explores the model inside the metallic phase and shows our description of the disordered local Fermi liquid.  The full model connecting both solutions is discussed in Sec.~\ref{sec:2kondo} and we show how it smoothly connects the previous limits.  Sec.~\ref{sec:conc} concludes the paper.  Appendix~\ref{sec:details} discusses further details of the large-$\mathcal{N}$ solution, Appendix~\ref{sec:finite} shows the robustness of our conclusions with respect to finite-size effects,  and Appendix~\ref {sec:choice} explores further the details of the full model.

%%%%%%%%%%%

\section{\label{sec:model} Model and large-$\mathcal{N}$ solution}

A possible minimal microscopic model for doped semiconductors consists of a single-band Hubbard model with $N_\mathrm{imp}$ dopant sites randomly distributed in the continuum of a cubic volume $L^3$ \cite{milovanovic89,  bhatt92}.  The hopping amplitude between sites $i$ and $j$ is $t_{ij}=t_0 e^{-r_{ij}/a}$, where $r_{ij}$ is the inter-site distance, $t_0=1$ sets the energy scale, and $a$ is the effective Bohr radius, which sets the distance scale for the impurity electrons.   Because of the very broad distribution of hoppings, see the inset of Fig.~\ref{fig:RS}(a) below, the \ updated {average hopping value} of a given site can be much smaller than $t_0$~\cite{milovanovic89, bhatt92}.  We disregard additional corrections to $t_{ij}$ arising from anisotropy \citep{herring64,  koiller02} and other effects, as the exponential behavior captures the leading contribution in the highly disordered diluted regime.  Unless otherwise stated, we study dopant concentrations in the range $ n=0.002-0.040$ and system sizes in the interval $L=20-50$, always at half-filling in the conduction-electron band.  We perform around $10^{3}-10^{4}$ realizations of disorder (dopant configurations) for each combination of $n$ and $L$, assuming periodic boundary conditions.  

%%%%%%%%%%%

\subsection{Non-interacting limit}

In the non-interacting limit ($U=0$, with $U$ being the on-site Coulomb repulsion),  the Hamiltonian reads:
\begin{eqnarray}
\mathcal{H}_{0} & = & -\sum_{ij,\sigma}t_{ij}c_{i\sigma}^{\dagger}c_{j\sigma},\label{eq:H0}
\end{eqnarray}
where $c_{i\sigma}^{\dagger}\left(c_{i\sigma}\right)$ creates (destroys) an electron at site $i$ with spin projection $\sigma$ and $t_{ij}$ are the hopping matrix elements between sites $i$ and $j$.  

\begin{figure}[t]
\begin{centering}
\includegraphics[width=1\columnwidth]{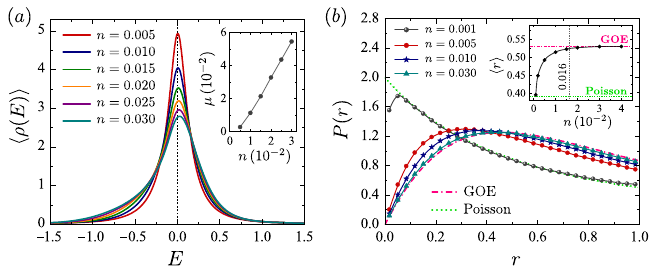}
\par\end{centering}
\caption{\label{fig:nonint}(a) Averaged density of states of the non-interacting Hamiltonian $\mathcal{H}_0$ in Eq.~\ref{eq:H0} as a function of energy $E$ for several values of dopant concentrations $n$. (Inset) Chemical potential $\mu$ as a function of $n$ for half-filling.  (b) Distribution of the level spacing $r$, $P(r)$, for several values of $n$.  Inset: average value of $r$ as a function of dopant concentration. The vertical dashed line marks the critical concentration $n_c=0.016$.  We considered $L=30$. }
\end{figure}

The averaged density of states (DOS) for the model at different impurity configurations is shown in Fig.~\ref{fig:nonint}(a).  The bandwidth covers the range $ -1.5 \lesssim E/t_0 \lesssim 1.5$, and it is nearly particle-hole symmetric, despite the existence of odd loops in the system because of the extended nature of the hopping. Therefore, for simplicity, we will chiefly consider $E=0$, which approximately describes a half-filled conduction band for all $n$ [with chemical potential $\mu$ close to zero, see the inset of Fig.~\ref{fig:nonint}(a)]. The DOS curve is featureless, displaying a slight increase in bandwidth as $n$ increases. 

To characterize the spatial extent of the bulk states and probe the MIT, we analyze the level statistics of the $\mathcal{H}_0$ spectrum. Specifically,  we diagonalize Eq.~\ref{eq:H0} for $L=30$ and several dopant concentrations, averaging over many disorder realizations for each $n$.  In each case, we considered an energy window around $E=0$ containing $1/16$ of the total number of states and evaluated the average level spacing ratio \cite{oganesyan07}
\begin{eqnarray}
r_{\nu}=\frac{\mbox{min}(\delta_{\nu},\delta_{\nu+1})}{\mbox{max}(\delta_{\nu},\delta_{\nu+1})}, \label{rstat}
\end{eqnarray}
where $\delta_\nu = E_{\nu+1} - E_\nu \ge0$ denotes the adjacent level spacing between consecutive eigenvalues $E_\nu$ of the matrix $\mathcal{H}_0$ in a given impurity configuration. The ratio $r_\nu$ is used to analyze spectral statistics.  For delocalized states, the distribution $P\left(r\right)$ follows the Gaussian orthogonal ensemble (GOE) because of their level repulsion,  with  $P(r) \sim r$,  for small $r$, and $\braket{r}_\mathrm{GOE}\approx 0.53$ \citep{atas13}. In contrast, strongly localized states show no level correlation and exhibit Poisson statistics, with $P(r) \to 2$ at small $r$,  and $\braket{r}_\mathrm{Poisson}\approx0.39$.  As shown in Fig.~\ref{fig:nonint}(b), both regimes across the Anderson MIT are successfully captured by the distribution $P(r)$.  We leave a more detailed finite-size scaling study of the transition for a future investigation and take $n_c=0.016$ as our critical dopant concentration following Refs.~\cite{root88,krich11}.  This critical density agrees remarkably well with experimental observations in conventional doped semiconductors~\cite{mott_mit}.  The established critical exponent,  $\nu=1.58\left(3\right)$ \citep{slevin99,vlad03,makros06,rodriguez10},  however,  does not agree with the experimental values \citep{paalanen91,lohneysen11}, pointing to the importance of correlations. 

%%%%%%%%%%%

\subsection{Effective model}

Despite its simplicity, the Hubbard model remains challenging to solve, especially in the presence of disorder~\cite{milovanovic89,denteneer99,heidarian04,statdmft}.  To overcome this difficulty,  we follow the phenomenological two-fluid model \cite{paalanen88,  paalanen91,  milovanovic89,  bhatt92},  and assume that the dopant electrons either form a local moment or become mobile,  contributing to the impurity band.  The localized spins will dominate the thermodynamic response,  whereas the moving electrons dominate the transport.  Moreover, experiments show that the fraction of dopant sites that form localized spins is no more than $10\%$ near the MIT, suggesting that the diluted-impurity limit is a reasonable starting point.
%We thus propose an effective Kondo-like low-energy model,  where diluted pairs of impurities,  corresponding to the local moments,  are coupled to a non-interacting disordered bath formed by the mobile electrons. This extends the original formulation, as we allow the local moments either to be Kondo-screened or to form random singlets.

To further rationalize this construction, consider a configuration in which two impurities are widely separated from all the others. Because these isolated sites are singly occupied and weakly coupled to their surroundings, each behaves as a free local moment.  This mimics the presence of diluted impurity pairs in the system.  By contrast, we consider the remaining $N_\mathrm{imp}-2$ sites well connected and behaving as conduction electrons. Free local moments at low temperatures naturally lead to a singular thermodynamic response, and we then pose the question: at which energy scale are these local moments effectively quenched via either the Kondo effect or a direct magnetic interaction?

To address the low-$T$ thermodynamics quantitatively,  we adopt the two-impurity Kondo-Heisenberg model as our minimal model, where a band of $c$-electrons hybridizes with localized spins located at positions $\boldsymbol{r}_{a\left(b\right)}$: 
\begin{align}
\mathcal{H} = \mathcal{H}_0 
&+ \sum_{ij\ell\alpha\beta} K_{ij}^\ell \left( c^\dagger_{i\alpha} \frac{\boldsymbol{\sigma}_{\alpha\beta}}{2} c_{j\beta} \right) \cdot \boldsymbol{S}_\ell + J_{ab} \, \boldsymbol{S}_a \cdot \boldsymbol{S}_b.
\label{eq:KH2}
\end{align}
Here, $\mathcal{H}_0$ describes the non-interacting conduction electron band [see Eq.~\ref{eq:H0}], $\boldsymbol{\sigma}$ is a vector containing the Pauli matrices, and $\boldsymbol{S}_{\ell}$ is the localized spin operator at the impurity sites $\ell =a,b$. The Kondo and Heisenberg couplings can be obtained via Schrieffer-Wolff transformations and are given by ~\cite{ong11}: $K_{ij}^{\ell}\approx 8t_{i\ell}t_{\ell j}/U$ and $J_{ab}=4t_{ab}^2/U$.  See also Appendix~\ref{sec:details}. The non-local nature of the Kondo coupling is an important aspect of the model, as it naturally incorporates a dependence on the impurity concentration $n$.

Within the effective description, local-moment formation is treated phenomenologically, while the subsequent competition between singlet formation and Kondo screening is addressed explicitly.  For simplicity, we set $U/t_0=1$ throughout this work ~\cite{thomas81,  milovanovic89,bhatt92}. 

%The degree of isolation of a given site will be naturally controlled by the strength of the Heisenberg and Kondo couplings. 

\subsection{Large-$\mathcal{N}$ solution}

To proceed, we solve the Kondo-Heisenberg model in Eq.~\ref{eq:KH2} within a large-$\mathcal{N}$ mean-field theory \citep{jones89}. We enlarge the local-moment symmetry to SU$\left(\mathcal{N}\right)$ and the localized spins are represented in terms of auxiliary fermions,  $S_{\ell}^{\alpha \beta}=f^{\dagger}_{\ell \alpha}f_{\ell \beta} - \delta_{\alpha,\beta}/2$,  
subject to an on-site constraint, $\sum_{\alpha=1}^{\mathcal{N}}f_{\ell \alpha}^{\dagger}f_{\ell \alpha}=\mathcal{N}/2$, where $f_{\ell \alpha}^{\dagger}\left(f_{\ell \alpha}\right)$ is the fermionic creation (destruction) operator and $\delta_{\alpha,\beta}$ the Kronecker delta.  In this fermionic language, we introduce two Hubbard-Stratonovich fields $b_{\ell}$, conjugate to $\sum_{i\alpha}t_{i\ell}f_{\ell \alpha}^{\dagger}c_{i \alpha}$,  and $\Delta_{ab}=\Delta$, conjugate to $\sum_{\alpha}f_{b \alpha}^{\dagger}f_{a\alpha}$.  The local constraint is enforced via Lagrange multipliers $\lambda_{\ell}$. The Hamiltonian corresponding to Eq.~\ref{eq:KH2} becomes extensive in $\mathcal{N}$ by rescaling the coupling constants as $K_{ij}^{\ell}/2\to K_{ij}^{\ell}/\mathcal{N}$, $J_{ab}/2\to J_{ab}/\mathcal{N}$.  

In the $\mathcal{N}\to\infty$ limit, the physics is controlled by a saddle point where the Hubbard-Stratonovich fields condense \citep{jones89,senthil08,hackl08,coleman15}.  We thus obtain the effective mean-field free energy for the two local moments and the corresponding self-consistency equations for the variational parameters $\lambda_{\ell}$,  $b_{\ell}$,  and $\Delta$. As discussed in Appendix~\ref{sec:details},  we solve these equations in the limit $b_{\ell}\to 0$ and $\lambda_{\ell} \to 0$.  This corresponds to approaching the instability of the high-temperature local-moment phase toward Kondo screening. Therefore, this procedure determines the onset scale for Kondo screening, namely the pair Kondo temperature $T_K^{\mathrm{pair}}$, analogously to the standard Nagaoka-Suhl treatment of the single-impurity Kondo problem ~\cite{suppl,Hewson_kondo}.  Our objective is not to obtain the complete low-temperature heavy Fermi-liquid solution, but rather to uncover the energy scale at which local moments become quenched by the conduction bath and to compare this scale against random-singlet formation,  fixed by the amplitude of the singlet gap $|\Delta|$.

The resulting mean-field equations dictate that (i) $\Delta \ne  0$ only if $J_{ab} > 4T_{K,\mathrm{max}}^0$, where $T_{K,\mathrm{max}}^0=T_K^\mathrm{pair}(\Delta=0)=\max(T_{K,a}^0,T_{K,b}^0)$ is the largest single-impurity Kondo temperature $(T_{K,\ell}^0)$ in the pair, and (ii) $|\Delta|=J_{ab}/2$ when $T_{K}^\mathrm{pair}=0$ ~\citep{dong21}.  This translates into three possible solutions: $(\mathrm{S}1)$ the random-singlet solution, $|\Delta|=J_{ab}/2$ and $T_{K}^\mathrm{pair}=0$, corresponding to decoupling the local moments from the conduction electrons but coupling them to each other; $(\mathrm{S}2)$ the single-impurity Kondo solution, $|\Delta| = 0$ and $T_{K}^\mathrm{pair}>0$, where the local moments are independently screened below their corresponding $T_{K,\ell}^0$; and $(\mathrm{S}3)$ the coexistence solution where both $|\Delta| $ and $T_{K}^\mathrm{pair}$ are non-zero.  In general, we find that solutions $(\mathrm{S}1)$ and $(\mathrm{S}2)$ are separated by a discontinuous transition as a function of an external parameter, for instance, $J_{ab}$. Moving beyond large-$\mathcal{N}$, one generally expects a crossover \citep{fye94,affleck95}. Moreover, we find that solution $(\mathrm{S}3)$ occurs in a small range of the parameter space and is drastically suppressed as the impurity separation and/or asymmetry in the Kondo couplings increases~\cite{suppl}.  We now discuss these solutions in detail, highlighting the nontrivial effects introduced by the broad disorder distribution.  

%%%%%%%%%%%
 
\section{\label{sec:RS}Random-singlet formation}

Let us discuss the non-Kondo solution,  $|\Delta|=J_{ab}/2$ and $T_{K}^\mathrm{pair}=0$,  in which we effectively decouple the two local moments from the conduction electrons. Motivated by the Bhatt-Lee theory of random-singlet formation \cite{bhatt_lee82}, which successfully describes the low-energy physics of Si:P deep inside the insulating phase, we randomly place $N_\mathrm{imp}$ dopant sites in a cubic volume, and assume that they all form local moments with $T_{K,\ell}^0=0$ (no Kondo effect). We treat this setup as an ensemble of $N_\mathrm{imp}/2$ independent two-site problems for each dopant configuration. To determine the corresponding spin pairs of a given sample,  we employ the hierarchical scheme introduced in Ref.~\citep{zhou09}. First, we identify the pair with the strongest coupling $J_{ab}$, which corresponds to the shortest distance between spins. Next, we eliminate (decimate) this pair from the system by forming a singlet. Finally, we iteratively repeat these first two steps until all $N_\mathrm{imp}/2$ pairs $a$ and $b$ are formed.  Importantly,  this decimation approach proceeds without any additional renormalization steps. 

%This approach is motivated by the original random-singlet formation \cite{bhatt_lee82}.

\begin{figure}[t]
\begin{centering}
\includegraphics[width=1\columnwidth]{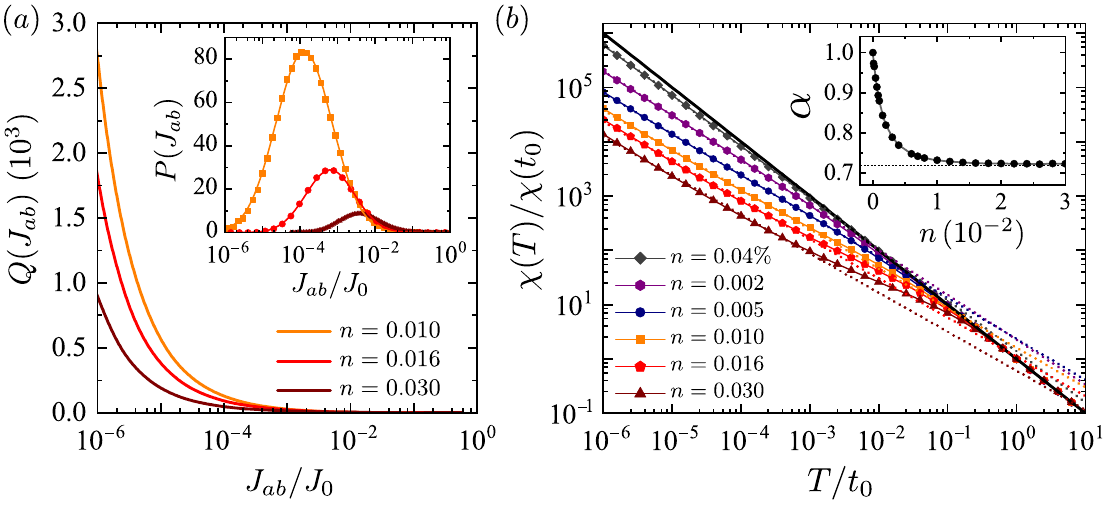}
\par\end{centering}
\caption{\label{fig:RS}(a) Normalized distribution of the decimated exchange couplings $Q\left(J_{ab}\right)$ for three different values of impurity concentration $n$, with $J_0=4t_0^2/U$.  Inset: Distribution of the nearest-neighbor couplings $P\left(J_{ab}\right)$ for the same $n$; (b) Magnetic spin susceptibility $\chi\left(T\right)$ as a function of the temperature $T$ for several $n$ on a log-log scale. The dotted lines are fits to a power law, $\chi\left(T\right) \sim T^{-\alpha}$ with the concentration dependence of the exponent $\alpha$ shown in the inset. The black line is the Curie law $\chi\left(T\right) = T^{-1}$. We considered $L=30$.}
\end{figure}

For a given temperature $T$, we say that spin pairs with $|\Delta|=J_{ab}/2>T$ are frozen in a spin-singlet state and do not contribute to the susceptibility, whereas those with $|\Delta|<T$ are essentially free. The spin susceptibility is then $\chi\left(T\right)=\nf/T$, where $\nf$ is the density of free spins at a given temperature $T$. This density is given by $\nf =n \int_{0}^{2T}Q\left(J\right)dJ$, where $Q\left(J_{ab}\right)$ is the distribution of the decimated couplings, shown in Fig.~\ref{fig:RS}(a) for three different impurity concentrations.  We also show the distribution $P\left(J_{ab}\right)$ of the nearest-neighbor couplings for comparison. Because of the hierarchical decimation procedure,  pairs at progressively larger separations are formed as the energy decreases, and the distribution $Q\left(J_{ab}\right)$ inevitably becomes singular, whereas $P\left(J_{ab}\right)$, well-fitted by a log-normal distribution, is broad but finite.  This singular behavior of $Q\left(J_{ab}\right)$ highlights a broad distribution of energy scales and suggests that the interaction between pairs will have a subleading thermodynamic effect.

As shown in Fig.~\ref{fig:RS}(b), $\chi\left(T\right)$ inherits the singularity of $Q\left(J_{ab}\right)$ at low $T$.  To quantify this behavior,  we fit the data as $\chi\left(T\right) \sim T^{-\alpha}$, and we observe that this power law captures the data well at sufficiently low temperatures,  with the exponent $\alpha$ evolving sharply from the Curie value as $n\xrightarrow{}0$ to $\alpha \approx 0.7$ for $n \gtrsim 0.010$, in the temperature range we considered. For low dopant concentrations, the power-law behavior occurs below $T/t_0 \lesssim 10^{-2}$. As discussed in Ref.~\citep{suppl}, the exact form of the susceptibility can be obtained for this problem \cite{zhou09}. Still, we find it helpful to fit $\chi\left(T\right) \sim T^{-\alpha}$ because: (i) it recovers the original Bhatt-Lee solution, reinforcing that the correlated hierarchical temperature dependence captures the random-singlet formation, and (ii) it connects naturally with the single-impurity Kondo solution, as we show next.

% At extremely low concentrations, all spins are essentially free and $\chi(T)$ must converge to the Curie law, $\alpha \to 1.$%

%%%%%%%%%%%

\section{\label{sec:1kondo}Single-impurity Kondo problem}

\begin{figure}[t]
\begin{centering}
\includegraphics[width=1\columnwidth]{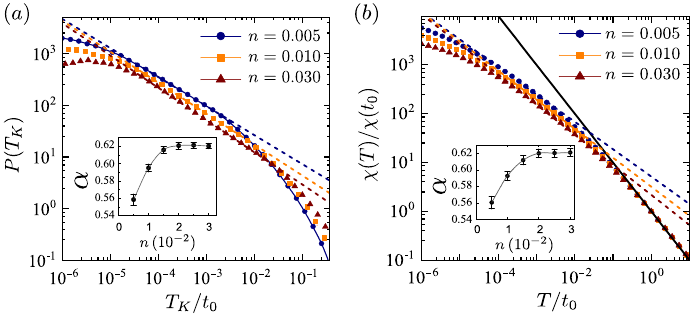}
\par\end{centering}
\caption{\label{fig:kondo1}Distribution of single-impurity Kondo temperatures $P\left(T_{K}\right)$ for different impurity concentrations $n$ on a log-log scale. The dashed lines are power-law fits of the form $T_{K}^{-\alpha}$. We have $\alpha\approx0.6$ for all $n$, see inset; (b) Susceptibility $\chi\left(T\right)$ as a function of $T$ on a log-log scale. The dashed lines are fits $\chi\left(T\right)\propto T^{-\alpha}$
with the exponent $\alpha \approx 0.6$ displayed in the inset. The full black line is the Curie law. We considered $L=30$.}
\end{figure}

We now discuss the single-impurity Kondo solution, $|\Delta|=0$ and  $T_K^\mathrm{pair}>0$. Here, the local moments are independently Kondo screened. Again, we distribute the local moments at the impurity positions, but now treat the system as an ensemble of $N_\mathrm{imp}$ single-impurity Kondo problems. Within our model, this means that we select a given site and assume it forms a local moment. The remaining $N_{\mathrm{imp}}-1$ sites then constitute an inhomogeneous non-interacting electronic bath. The disordered nature of the problem gives rise to a broad distribution of local single-impurity Kondo temperatures $P\left(T_{K}\right)$ \cite{vlad92a,miranda01,nfl_review05,cornaglia06,kaul07,kettemann09,kettemann12,miranda14,andrade15}, which is well described by $P(T_{K}) \propto T_{K}^{-\alpha}$ at low $T_{K}$, see Fig.~\ref{fig:kondo1}(a). Once $T_{K}$ falls below the typical mean level spacing $\delta \propto L^{-3}$ of the conducting band, we lose spectral resolution, and $P\left(T_{K}\right)$ departs from its power-law behavior.  We verified that, as $L$ increases, this deviation point is indeed pushed to lower energy scales; see Appendix~\ref{sec:finite}. Furthermore, we find that the exponent $\alpha$ depends weakly on the impurity concentration $n$, with $\alpha \approx 0.6$; see Fig.~\ref{fig:kondo1}(a),  although the non-local Kondo couplings vary with $n$. Previous investigations of the single-site Kondo effect in disordered baths also report this peculiar weak dependence of the exponent $\alpha$ on disorder \citep{cornaglia06,kettemann09,kettemann12,miranda14}. In contrast, a non-universal power-law exponent is expected within the Electronic Griffiths Phase scenario of the disordered Kondo problem \citep{darko04,  nfl_review05}.  

To connect $P\left(T_{K}\right)$ with thermodynamic quantities,  we use the interpolation formula: $\chi\left(T\right)=\left\langle1/ \left(T+T_{K,\ell}^0\right)\right\rangle$ \citep{Hewson_kondo}, see Fig.~\ref{fig:kondo1}(b), where $\left< \cdots \right>$ denotes average over sites and dopant configurations. This expression gives the correct behavior at both high and low temperatures, with the crossover from free local moments to the Kondo regime occurring around $T/t_{0} \approx 10^{-2}$. Below this temperature, we get $\chi\left(T\right)\propto T^{-\alpha}$, with the same exponent $\alpha$ of $P\left(T_{K}\right)$, reflecting the disorder-induced broad distribution of $T_{K}$.

%Despite the fact that the exponent $\alpha$ decreases at small $n$,  naively indicating a less singular behavior for large disorder.  However,  the fraction of free spins,  the sites that exhibit $T_K=0$,  is largely dominant in this regime,  overwhelming the singular distribution of Kondo temperatures.  

Besides sites with a finite Kondo temperature, we also find a fraction of free spins $f(n,L)$, i.e.,  local moments with $T_{K,\ell}^0=0$,  which depends on both $n$ and $L$. From the mean-field equations, the condition for a local moment at site $d$ having a finite Kondo temperature reads $\sum_{ij\nu}\braket{i|\nu}\braket{\nu|j}K_{ij}^d/|E_\nu| > 2$,  where $\ket{\nu}$ and $E_\nu$ are, respectively, the eigenvectors and eigenvalues of $\mathcal{H}_0$. Assume, then, that we are on the insulating side of the MIT and that the wavefunction $\braket{i|\nu}$ is Anderson-localized. Because the couplings $K_{ij}^d$ are exponentially suppressed by disorder, we conclude that the inequality may be violated. This justifies the existence of a finite fraction of spins that remain Kondo-unscreened down to $T \to 0$. As we discuss in Appendix~\ref{sec:finite}, our numerical results are consistent with this fraction vanishing in the metal and remaining finite in the insulator; see Fig.~\ref{fig:fss}.

The central result of this work is that both hierarchical random-singlet formation and disorder-induced Kondo screening independently generate singular thermodynamic responses. In the next section, we show that both solutions are smoothly connected in the experimentally relevant doping regime.

%We have thus uncovered two mechanisms that generate a singular low--$T$ thermodynamic response in our system, $\chi\left(T\right)\propto T^{-\alpha}$, namely the random-singlet formation -- due to the hierarchical decimation of spin singlets -- and the broad distribution of local Kondo temperatures -- due to the random conduction bath.  Although both ultimately stem from strong disorder, their physical origins are different.  We will now discuss their combined effects.

%%%%%%%%%%%

\section{\label{sec:2kondo}Two-impurity Kondo problem}

We now study an ensemble of two-impurity Kondo problems as an effective model for the thermodynamic behavior of strongly disordered systems. We start with the single-impurity Kondo problem.  First, we solve the mean-field equations and determine $T_{K,\ell}^0$ at all sites. Second, we split the solutions into two populations. Sites with finite $T_{K,\ell}^0$ form initially the Kondo population,  whereas those with $T_{K,\ell}^{0}=0$ comprise the random-singlet  population.  Third, we construct an ensemble of two-impurity Kondo problems for the Kondo  species employing our hierarchical procedure. For each of these pairs, we compare two energy scales: the intersite Heisenberg exchange $J_{ab}$ and the largest single-impurity Kondo temperature $T_{K,\mathrm{max}}^0=\max{\left(T_{K,a}^0,T_{K,b}^0\right)}$. For  $J_{ab} \le 4T_{K,\mathrm{max}}^\mathrm{0}$, $\Delta=0$ and both sites retain their single-impurity Kondo temperature.  If, on the other hand, $J_{ab} > 4T_{K,\mathrm{max}}^\mathrm{0}$, we solve Eqs. \ref{eq:self1} and \ref{eq:self2}. We find that, in almost all cases, $|\Delta|=J_{ab}/2$ and $T_{K}^\mathrm{pair}=0$, implying that we must transfer these pairs from the Kondo  population to the random-singlet  population. The remaining pairs stay in the Kondo  population with a slightly modified local Kondo temperature.  See Appendix~\ref{sec:choice} for further details.  

Fig.~\ref{fig:kondo2}(a) shows the evolution of each species fraction as a function of $n$ at $T=0$. As expected, the random-singlet solution dominates deep into the insulator, whereas the Kondo solution is enhanced in the metal, indicating the formation of a random local Fermi liquid. Interestingly, the fraction of each population evolves smoothly with doping and becomes equal near $n_c$.  Physically, this means that the local moments are more likely to form a singlet deep in the insulator. In contrast, they are typically Kondo-screened by the metallic bath as we cross the MIT, in a disordered version of Doniach's competition between magnetism and the Kondo effect~\cite{coleman15,sachdev2023}. Once we identify the sites for each population, we determine the thermodynamic behavior of each species separately, following the same procedure as before, and then add their contributions. As shown in Fig.~\ref{fig:kondo2}(b), the total susceptibility $\chi(T)$ exhibits a power-law divergence at low temperatures, $\chi(T) \sim T^{-\alpha}$,  for all $n$,  with the exponent $\alpha$ decreasing continuously with $n$ within the insulating phase and converging to a constant value $\alpha \approx 0.6$ near and above $n_c$, Fig.~\ref{fig:kondo2}(b),  as a result of this competition between random-singlet formation and Kondo screening.  This trend agrees with experimental work on heavily doped semiconductors ~\citep{andres81, paalanen91, nfl_review05, lohneysen11, vlad12}.
%showing that the exponent $\alpha$ beautifully interpolates between two regimes: the random-singlet phase at low impurity concentrations and the inhomogeneous local-Fermi liquid beyond the MIT, where the conduction electrons screen the magnetic impurities.

%%%%%%%%%%%

\section{\label{sec:conc}Conclusions}

We investigated the two-site Kondo-Heisenberg model in a strongly disordered bath, Eq.~\eqref{eq:KH2}, as an effective low-energy model for the singular low-temperature thermodynamic behavior observed in doped semiconductors. Because disorder is the dominant energy scale in this system, we expect spin-spin correlations beyond pairs to be subleading. Using a consistent large-$\mathcal {N}$ approach, combined with a hierarchical decimation scheme, we find two solutions.  One is the non-Kondo solution, more pronounced on the insulating side, where the spins decouple from the conduction bath and form random singlets. The second solution is the Kondo solution, which describes the quenching of these local moments by conduction electrons, leading to an inhomogeneous Fermi liquid.  These solutions coexist in a given sample, with their relative fractions changing with the dopant concentration.

Because the bare energy scales of the system are broadly distributed owing the random position of the dopants, both solutions lead to a singular thermodynamic behavior, which can be effectively described as $\chi(T) \sim T^{-\alpha}$ for all concentrations,  indicating that unlike the homogeneous system,  thermodynamic quantities display no sharp distinction between the two solutions.  The power-law exponent $\alpha$ is a smooth function of the model parameters, even across the MIT, and saturates in the metallic regime, with $\alpha \approx 0.6$ in our model.  We expect more accurate approaches to improve the results quantitatively, especially at finite $T$ \citep{miranda14}.  However, they should not substantially alter our conclusions, since the system's strongly disordered nature is the primary driving force behind this behavior \cite{kettemann23}.

% \update{Since the present work is focused solely on low-energy spin thermodynamics rather than on a complete microscopic theory of the interacting metal-insulator transition itself,} 

As our next step, we would like to solve the disordered Hubbard model ~\citep{milovanovic89,heidarian04,byczuk05,carol09,amd09,mard15}, thereby dynamically establishing the formation of local moments and going beyond the phenomenological description of the two-fluid behavior by self-consistently coupling mobile electrons and local moments.   The existence of unscreened local moments inside the metal,  as suggested experimentally,  inevitably causes a violation of Luttinger's theorem since a fraction of the electrons would not contribute to the mobile electron Fermi sea \citep{coleman15, burdin00,  oshikawa00, senthil03,  lechermann07,  lee16,  najera17, sachdev2023, drechsler25}, implying non-Fermi-liquid behavior \citep{nfl_review05} and a singular thermodynamic response.  Moreover, a finite density of such ``deconfined'' local moments at the MIT should fundamentally alter its universality class, with important implications for the critical exponents and promising a significant advance in our understanding of the MIT in doped semiconductors. 

%.  Within this picture, one might envision a novel perspective on the entire phase diagram

%%%%%%%%%%%
\begin{acknowledgments}
We acknowledge helpful discussions with Vitor Dantas, Tobias Meng, Carsten Timm, and Matthias Vojta.  We acknowledge support by FAPESP (Brazil),  Grants No.  2019/17645-0,  2021/06629-4,  2022/15453-0,  and 2023/06094-9.  ECA and EM were also supported by CNPq (Brazil), Grant Nos. 302823/2022-0 and 309584/2021-3, respectively. Work in Florida (V.D.) was supported by
the NSF Grant No. DMR-2409911, and the National High
Magnetic Field Laboratory through the NSF Cooperative
Agreement No. DMR-2128556 and the State of Florida.
\end{acknowledgments}

%%%%%%%%%%%

\section*{DATA AVAILABILITY}

The data supporting the findings of this article are openly available \citep{Rabelo2026data}.

\appendix

\section{\label{sec:details}Details of the large-$\mathcal{N}$ solution}

Our primary motivation is to understand the singular low-temperature thermodynamic response in doped semiconductors near the MIT.  In Si:P, for instance, each phosphorus ion forms a bound state approximately $45$ meV below the silicon conduction band for $n \to 0$. Adding an extra electron to this bound state costs an on-site Coulomb repulsion $U$, which is very close to the binding energy~\citep{thomas81}. This suggests a local-moment picture on the insulating side.  

As the impurity density increases and we enter the metallic phase, one expects the quenching of local moments and the formation of a disordered Fermi liquid. However, strong disorder allows a fraction of local moments to survive, even inside the metal, at experimentally relevant temperatures \citep{milovanovic89,byczuk05,carol09,amd09}. These general observations led us to propose Eq.~\ref{eq:KH2} as our minimal microscopic model. This model combines two sources of singular thermodynamic behavior for local moments in contact with a random bath: the random-singlet formation and Kondo screening with a broad distribution of Kondo temperatures. 

To further motivate our effective model, let us consider $N_\mathrm{imp}$ dopant sites randomly placed in the continuum of a cubic volume. We assume that two arbitrary sites at positions $\bm{r}_a$ and $\bm{r}_b$ form localized moments, while the remaining $N_\mathrm{imp}-2$ sites constitute a conduction band.  As discussed in the main text, one might rationalize this local-moment formation by imagining isolated singly occupied sites weakly coupled to their surroundings. In contrast, the remaining sites are well-connected and can be described as conduction electrons. The corresponding two-impurity Anderson model  reads
\begin{align}
\mathcal{H}_{\mathrm{A}} &=
 -\sum_{ij\ne \ell,\sigma} t_{ij}c_{i\sigma}^\dagger c_{j\sigma} - \mu\sum_{i\ne \ell,\sigma} c_{i\sigma}^\dagger c_{i\sigma}+ U \sum_{\ell} n_{\ell\uparrow} n_{\ell\downarrow} 
 \nonumber\\
&\;
 -\sum_{i\ne \ell,\sigma} \left( t_{i\ell}c_{i\sigma}^\dagger f_{\ell\sigma}
     +\mathrm{h.c.}\right)+\sum_{\ell,\sigma}\left(\epsilon_{d\ell}-\mu \right)n_{\ell\sigma}.
\label{anderson-hubbard}
\end{align}
where $i$ and $j$ run over conduction electron sites, $\ell=a,b$ labels the local moment sites, $c_{i\sigma}^\dagger$ ($f_{\ell\sigma}^\dagger$) is the creation operator of a conduction electron (localized moment) at site $i$ ($\ell$) with spin projection $\sigma$, $n_{\ell \sigma}=f_{\ell\sigma}^\dagger f_{\ell \sigma}$ is the occupation number operator at site $\ell$, $\epsilon_{d\ell}$ and $U$ are the on-site energies and local Hubbard repulsion for the local moment sites, respectively. We assume that the conduction bath possesses an approximate particle-hole symmetry and set $\epsilon_{d\ell}-\mu \approx -U/2$, see Fig. ~\ref{fig:nonint}(a).  To derive Eq. \ref{eq:KH2}, we (i) apply a Schrieffer--Wolff (SW) transformation in Eq.~\eqref{anderson-hubbard} up to second order in the hopping amplitudes, thereby projecting out empty and doubly occupied configurations on the local moment sites, and (ii) introduce a direct Heisenberg coupling ($J_{ab}$) between them.  Notice that we have a very broad distribution of hoppings,  which gives rise to sites that are weakly coupled to the bath and thus are prone to form local moments; see the inset of Fig.~\ref{fig:RS}.  The SW transformation also generates an indirect RKKY interaction at fourth order \cite{ong11}, which is suppressed in the large-$\mathcal{N}$ limit we present next $(J_\mathrm{RKKY}\sim 1/\mathcal{N}^2)$.  

To solve the Kondo-Heisenberg model, Eq.~\eqref{eq:KH2} of the main text, we employ the large-$\mathcal{N}$ static mean-field theory of Ref.~\citep{jones89},  also presented in the main text.  Within this approach,  we obtain the following effective mean-field free energy for the two local moments \citep{jones89,senthil08,hackl08,coleman15,dong21}
\begin{equation}
\begin{aligned}
\frac{F_{\mathrm{mf}}}{\mathcal{N}}
   = -T\sum_{n}\ln \det \mathcal{O}_n+\sum_{\ell}\frac{|b_\ell|^{2}}{g}
      +\frac{|\Delta|^2}{J_{ab}}
      -\frac{{1}}{2}\sum_{\ell}\lambda_{\ell},
\end{aligned}
\label{eq:F_MF_HK2}
\end{equation}
where the matrix $\mathcal{O}_n$ reads
\begin{equation}
    \mathcal{O}_n=\begin{pmatrix}
        -i\omega_{n}+\lambda_{a}+|b_a|^2\phi_{aa}
          & \Delta+b_{a}^{\star}b_{b}\,\phi_{ab} \\[2pt]
        \Delta^{\star}+b_{a}b_{b}^{\star}\, \phi_{ba}
          & -i\omega_{n}+\lambda_{b}+|b_b|^{2}\phi_{bb} \nonumber
      \end{pmatrix}.
\end{equation}
Here, $\omega_{n}$ are fermionic Matsubara frequencies, $g=8/U$,  and the non-local hybridization function $\phi_{\ell \ell'}(i\omega_n)$ reads
\begin{equation*}
\phi_{\ell \ell'}(i\omega_n)
= \sum_{ij\nu} t_{i\ell } t_{\ell' j}
  \frac{\langle i|\nu\rangle \langle \nu|j\rangle}{i\omega_n - E_\nu},
\end{equation*}
with $E_\nu$ and $\ket{\nu}$ denoting, respectively, the eigenvalues and eigenvectors of the conduction bath Hamiltonian $\mathcal{H}_0$.  The mean-field equations are obtained by minimizing the free energy in Eq. \eqref{eq:F_MF_HK2} with respect to the variational quantities $\{ \lambda_a,\lambda_b,b_a,b_a^{\star},b_b,b_b^{\star},\Delta,\Delta^{\star}\}$.  Because of the local freedom $f_{\ell\sigma}\xrightarrow{}e^{i\theta_\ell}f_{\ell\sigma}$, $b_\ell \xrightarrow{} e^{-i\theta_\ell}b_\ell$, and $\Delta \xrightarrow{}e^{i(\theta_a-\theta_b)}\Delta$, we can work in a gauge where $b_a$ and $b_b$ are real, while $\Delta$ is complex \cite{jones89,sachdev2023,coleman15,dong21}.  At high temperatures,  the saddle point yields $\bm{b}=(b_a,b_b)^T=\bm{0}$. However, as the temperature falls below a characteristic scale $T_K^\mathrm{pair}$, Kondo screening sets in and the local moments begin to be quenched by the conduction electrons, indicating that $\bm{b}\ne \bm{0}$.  To simplify the problem,  we work in the limit $\bm{b} \to \bm{0}$, $T \to T_K^\mathrm{pair}$,  where we find $\lambda_a=\lambda_b=0$ and $\Delta =\Delta^{\star}$. 

The saddle point conditions then become
\begin{gather}
\tanh\!\big(|\Delta|/2T_K^{\mathrm{pair}}\big)
  = 2|\Delta|/J_{ab}, \label{eq:self1}\\
\mathcal{A}_a+\mathcal{A}_b+4g^{-1}
  = \sqrt{(\mathcal{A}_a-\mathcal{A}_b)^2+4\mathcal{B}^2}.
  \label{eq:self2}
\end{gather}
where $\mathcal{A}_\ell=F_{\ell \ell}^-+F_\mathcal{\ell \ell}^+$, $\mathcal{B}=F_{ab}^{-}-F_{ab}^+$, and
\begin{align*}
    F_{\ell \ell'}^{\pm}(|\Delta|,T_K^{\mathrm{pair}})=\sum_{ij\nu}t_{i\ell }t_{\ell'j} \braket{i|\nu}\braket{\nu|j} \frac{f\left(E_\nu\right)-f\left(\mp|\Delta|\right)}{E_\nu\pm|\Delta|}.
\end{align*}
Here,  $f\left(x\right)$ is the Fermi-Dirac distribution at $T=T_K^\mathrm{pair}$.  There is an extra trivial equation, $f\left(|\Delta|\right) + f\left(-|\Delta|\right) = 1$, stating that the $f$-electron occupation is distributed among the two energy levels $\Delta$. Further details of the mean-field solution are presented in Ref.~\cite{suppl}.

\section{\label{sec:finite}Finite-size effects in the Kondo problem}

\begin{figure}[!t]
\begin{centering}
\includegraphics[width=1\columnwidth]{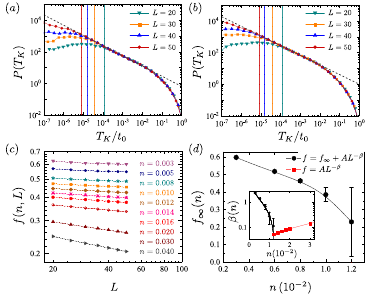}
\par\end{centering}
\caption{\label{fig:fss} $P(T_K)$ for several $L$ for (a) $n=0.010$ and (b) $n=0.020$. Vertical dashed lines mark the approximate finite-size cutoff $T_K^*\approx L^{-3}$, below which the distributions deviate from the power-law tail because of the finite-size gaps. (c) Fraction of spins that are not Kondo screened, $f(n,L)$,  as a function of the system size $L$ and several $n$ on a log-log scale. (d) Extrapolated fraction of free spins in the thermodynamic limit, $f_\infty(n)$, as a function of $n$. Values of $f_\infty(n)$ are obtained from finite-size scaling fits of $f(n,L) = f_\infty(n) + A(n)L^{-\beta(n)}$. Inset: Exponent $\beta$ as a function of $n$.}
\end{figure}
%Coefficient $A$ as a function of $n$. 

As discussed in the main text, the distribution of single-impurity Kondo temperatures has a power-law divergence at low temperatures, $P(T_K)\propto T_K^{-\alpha}$, that is truncated by finite-size effects for $T_K<T_K^*(L)$ where $T_K^*(L)$ is a size dependent cutoff scale of the order of the level spacing $\delta(L)\propto L^{-3}$.  In Fig.~\ref{fig:fss}(a-b), we show $P(T_K)$ for several $L$ in the insulating and metallic phases,  and we confirm that it displays a power-law regime for all $L$ over an appreciable window of Kondo temperatures, with no significant difference in the behavior inside the metal and insulator.   For $T_K<T_K^*(L)$,  the curves deviate from the power-law behavior,  indicating a saturation.  The scale $T_K^*(L)$ shifts systematically to lower energies as $L$ increases, but $P(T_K)$ displays no other prominent features, suggesting that the results for finite systems can capture the correct low-energy behavior in the experimentally relevant temperature range.

Apart from these finite-$T_K$ solutions, we also find a fraction $f(n,L)$ of sites that do not undergo Kondo screening, i.e., $T_{K,\ell}^0=0$, Fig.~\ref{fig:fss}(c).  We expect a finite fraction of spins that remain Kondo-unscreened down to $T \to 0$ inside the insulator.  On the metallic side,  we expect the Kondo effect to always occur in the thermodynamic limit.  To test this idea,  we fit the fraction of free spins as $f(n, L)=f_\infty(n)+A(n)L^{-\beta(n)}$.  As we show in Fig.~\ref{fig:fss}(d),  for $n \lesssim 0.012$,  the best fits require $f_\infty>0$,  consistent with a finite fraction of spins as $L \xrightarrow{}\infty$.  This fraction decreases with $n$ as the localization length of the conduction electrons increases.  Physically,  free spins arising in the insulator will not remain free down to $T \to 0$ owing to the random-singlet formation.  For $n \gtrsim0.012$,  the fit works better with $f_\infty=0$,  as the error bars in $f_\infty(n)$ become large, consistent with the general expectation that all spins should be screened in the metal.  

A similar trend appears in the fit for the other parameters.  For $n \gtrsim0.012$,  the amplitude $A(n)$ is essentially constant (not shown). The finite size scaling is governed by the exponent $\beta(n)$,  which increases with $n$,  indicating a drastic reduction in $f(n,L)$ as we move deeper into the metal; see the inset of Fig.~\ref{fig:fss}(d). Within error bars, all fit parameters at low and high $n$ match at $n=0.012$, and we take this concentration as the point at which the behavior changes.  This value is close to the conduction-band MIT.  Deeper into the insulator,  the exponent $\beta(n)$ increases rapidly as $n$ decreases,  implying that the fraction of free spins quickly reaches $f_\infty$.

\section{\label{sec:choice}Choice of basis}

Ultimately, we model our problem as an ensemble of $N_\mathrm{imp}/2$ two-impurity Kondo problems, assuming that all electrons in the sample can form a local moment. At low temperatures, each pair has two possible fates: either it is Kondo-screened by the conduction electrons, or it forms a random singlet. Both mechanisms give rise to a singular thermodynamic behavior, with the latter dominating in the insulating regime (low doping) and the former in the metallic regime (higher doping).

\begin{figure}[t]
\begin{centering}
\includegraphics[width=1\columnwidth]{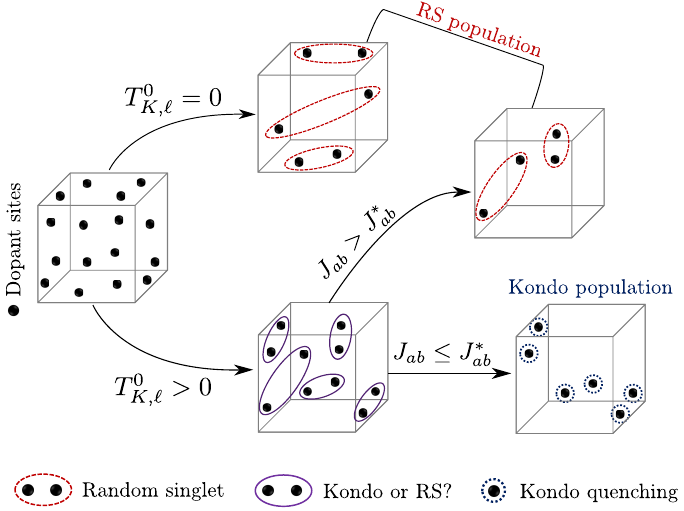}
\par\end{centering}
\caption{\label{fig:basis} Construction of the two-site basis employed in the main text to solve the disordered two-impurity Kondo problem within our large$-\mathcal{N}$ solution. We obtain a two-population solution: sites are either Kondo-screened by the conduction electrons or form a random singlet (RS). Here, $J^{*}_{ab} = 4T_{K}^{\rm{max}}$,  where $T_{K}^{\rm{max}}$ is the maximum single-impurity Kondo temperature of the two sites coupled by $J_{ab}$.}
\end{figure}

We explain our choice of basis in the main text; see Fig.~\ref{fig:basis}. We begin with the single-impurity Kondo problem, solve the mean-field equations, and determine $T_{K,\ell}^0$ at each dopant site. We then separate the solutions into two populations: sites with finite $T_{K,\ell}^{0}$ form the Kondo population, whereas sites with $T_{K,\ell}^{0}=0$ comprise the random-singlet population. After that, we construct an ensemble of two-impurity Kondo problems for the Kondo population employing the hierarchical decimation procedure of Ref.~\citep{zhou09}. For each of these pairs, we compare two energy scales: the intersite Heisenberg exchange $J_{ab}$ and the largest local single-impurity Kondo temperature $T_{K,\mathrm{max}}$. If  $J_{ab} \le J_{ab}^*$, where $J_{ab}^*=4T_{K,\mathrm{max}}^0$, $\Delta=0$ and both sites retain their single-impurity Kondo temperature.  If, on the other hand, $J_{ab} > J_{ab}^*$, Eqs.~\ref{eq:self1} and \ref{eq:self2} must be solved explicitly. We find that, in the overwhelming majority of these cases,  $\Delta=J_{ab}/2$ and $T_{K}^\mathrm{pair}=0$, implying that we must transfer these pairs from the Kondo to the random-singlet population. The remaining pairs stay in the Kondo population, with a slightly modified local Kondo temperature.  Incidentally, we remark that very few solutions ($\approx 3\%$ of their total number) fall within the coexistence interval, even for $n=0.030$, because of the strongly disordered nature of the problem.  As a result, we can essentially solve the full problem by just comparing energy scales of the single-impurity problem and random-singlet formation. 

Different basis choices are possible in general. Therefore, we also consider the random-singlet basis of the problem, which is built on the limit of decoupled impurity pairs.  We randomly place $n$ dopants in the system. We treat this setup as an ensemble of $N_\mathrm{imp}/2$ independent two-site problems (for each sample) and employ our hierarchical scheme.  First, we identify the spin pair with the strongest coupling $J_{ab}$, which corresponds to the shortest distance between spins. Next, we eliminate (decimate) this pair from the system by forming a singlet. Finally, we iteratively repeat these first two steps until all $N_\mathrm{imp}/2$ pairs $a$ and  $b$ are formed. 

\begin{figure}[!t]
\begin{centering}
\includegraphics[width=1\columnwidth]{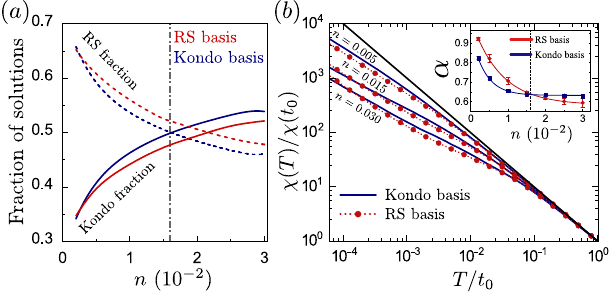}
\par\end{centering}
\caption{\label{fig:basis_comp} Comparison of the results for the different basis choices for solving the two-impurity problem. The Kondo basis and the random-singlet (RS) basis.  (a) Fraction of the two solutions, the Kondo solution (full curves) and the RS solution (dashed curves), as a function of the dopant concentration $n$ for the two bases.  (b) Susceptibility $\chi (T)$ as a function of the temperature $T$ for different  $n$.  We fit the result to a power law: $\chi (T) \sim T^{-\alpha}$.  (Inset) Power-law exponent $\alpha$ as a function of $n$.  The dashed vertical line marks the critical concentration $n_c$ for which the MIT in the conduction electrons takes place. }
\end{figure}

Once we define the random-singlet basis, we solve the self-consistency equations and compute $T_{K,\mathrm{max}}^0$ and $\Delta$ for all pairs.  Again, we obtain a two-species solution.  The results are presented in Fig.~\ref{fig:basis_comp}, where we compare the results for the two basis choices.  As we study the fraction of solutions in each population, we observe that the Kondo population overcomes the random-singlet one at a higher dopant concentration $n$ in the random-singlet basis than in the Kondo basis, because each basis favors the solution on which it was built.  As for the thermodynamic behavior, we see that qualitatively the two basis choices produce similar results. We obtain singular thermodynamics for all dopant concentrations that can be fitted to $\chi(T) \sim T^{-\alpha}$, with the exponent $\alpha$ decreasing with $n$. The random-singlet basis produces a more singular response in the insulator, as it favors the random-singlet formation, which has a larger $\alpha$ for $n \to 0$. In the metal, on the other hand, $\alpha$ is less singular than in the Kondo basis. Again, the tendency to form singlets is strong, but $J_{ab}$ is also enhanced,  which quenches the local moments at higher scales.

Ultimately, the strongly disordered nature of the problem dominates the physics, as the exchange coupling $J_{ab}$ is broadly distributed. Both the random-singlet formation and the Kondo quenching lead to a singular thermodynamic behavior,  $\chi(T) \sim T^{-\alpha}$,  persisting across the MIT, with the power-law exponent $\alpha$ a smooth function of the model parameters, independently of the choice of basis and details of the solution. We choose to work in the Kondo basis because it is deliberately constructed to bridge the random-singlet phase deep in the insulator with the inhomogeneous local Fermi-liquid that emerges at high dopant concentrations.


\begin{thebibliography}{66}%
\makeatletter
\providecommand \@ifxundefined [1]{%
 \@ifx{#1\undefined}
}%
\providecommand \@ifnum [1]{%
 \ifnum #1\expandafter \@firstoftwo
 \else \expandafter \@secondoftwo
 \fi
}%
\providecommand \@ifx [1]{%
 \ifx #1\expandafter \@firstoftwo
 \else \expandafter \@secondoftwo
 \fi
}%
\providecommand \natexlab [1]{#1}%
\providecommand \enquote  [1]{``#1''}%
\providecommand \bibnamefont  [1]{#1}%
\providecommand \bibfnamefont [1]{#1}%
\providecommand \citenamefont [1]{#1}%
\providecommand \href@noop [0]{\@secondoftwo}%
\providecommand \href [0]{\begingroup \@sanitize@url \@href}%
\providecommand \@href[1]{\@@startlink{#1}\@@href}%
\providecommand \@@href[1]{\endgroup#1\@@endlink}%
\providecommand \@sanitize@url [0]{\catcode `\\12\catcode `\$12\catcode `\&12\catcode `\#12\catcode `\^12\catcode `\_12\catcode `\%12\relax}%
\providecommand \@@startlink[1]{}%
\providecommand \@@endlink[0]{}%
\providecommand \url  [0]{\begingroup\@sanitize@url \@url }%
\providecommand \@url [1]{\endgroup\@href {#1}{\urlprefix }}%
\providecommand \urlprefix  [0]{URL }%
\providecommand \Eprint [0]{\href }%
\providecommand \doibase [0]{https://doi.org/}%
\providecommand \selectlanguage [0]{\@gobble}%
\providecommand \bibinfo  [0]{\@secondoftwo}%
\providecommand \bibfield  [0]{\@secondoftwo}%
\providecommand \translation [1]{[#1]}%
\providecommand \BibitemOpen [0]{}%
\providecommand \bibitemStop [0]{}%
\providecommand \bibitemNoStop [0]{.\EOS\space}%
\providecommand \EOS [0]{\spacefactor3000\relax}%
\providecommand \BibitemShut  [1]{\csname bibitem#1\endcsname}%
\let\auto@bib@innerbib\@empty
%</preamble>
\bibitem [{\citenamefont {Lee}\ and\ \citenamefont {Ramakrishnan}(1985)}]{lee_ramakrishnan}%
  \BibitemOpen
  \bibfield  {author} {\bibinfo {author} {\bibfnamefont {P.~A.}\ \bibnamefont {Lee}}\ and\ \bibinfo {author} {\bibfnamefont {T.~V.}\ \bibnamefont {Ramakrishnan}},\ }\bibfield  {title} {\bibinfo {title} {Disordered electronic systems},\ }\href {https://doi.org/10.1103/RevModPhys.57.287} {\bibfield  {journal} {\bibinfo  {journal} {Rev. Mod. Phys.}\ }\textbf {\bibinfo {volume} {57}},\ \bibinfo {pages} {287} (\bibinfo {year} {1985})}\BibitemShut {NoStop}%
\bibitem [{\citenamefont {Shklovskii}\ and\ \citenamefont {Efros}(1984)}]{shklovskii}%
  \BibitemOpen
  \bibfield  {author} {\bibinfo {author} {\bibfnamefont {B.~I.}\ \bibnamefont {Shklovskii}}\ and\ \bibinfo {author} {\bibfnamefont {A.~L.}\ \bibnamefont {Efros}},\ }\href@noop {} {\emph {\bibinfo {title} {Electronic Properties of Doped Semiconductors}}},\ Springer Series in Solid-State Sciences\ (\bibinfo  {publisher} {Springer},\ \bibinfo {year} {1984})\BibitemShut {NoStop}%
\bibitem [{\citenamefont {Paalanen}\ and\ \citenamefont {Bhatt}(1991)}]{paalanen91}%
  \BibitemOpen
  \bibfield  {author} {\bibinfo {author} {\bibfnamefont {M.}~\bibnamefont {Paalanen}}\ and\ \bibinfo {author} {\bibfnamefont {R.}~\bibnamefont {Bhatt}},\ }\bibfield  {title} {\bibinfo {title} {Transport and thermodynamic properties across the metal-insulator transition},\ }\href {https://doi.org/http://dx.doi.org/10.1016/0921-4526(91)90233-5} {\bibfield  {journal} {\bibinfo  {journal} {Physica B}\ }\textbf {\bibinfo {volume} {169}},\ \bibinfo {pages} {223 } (\bibinfo {year} {1991})}\BibitemShut {NoStop}%
\bibitem [{\citenamefont {Miranda}\ and\ \citenamefont {Dobrosavljevi{\'c}}(2005)}]{nfl_review05}%
  \BibitemOpen
  \bibfield  {author} {\bibinfo {author} {\bibfnamefont {E.}~\bibnamefont {Miranda}}\ and\ \bibinfo {author} {\bibfnamefont {V.}~\bibnamefont {Dobrosavljevi{\'c}}},\ }\bibfield  {title} {\bibinfo {title} {Disorder-driven non-fermi liquid behaviour of correlated electrons},\ }\href {https://doi.org/10.1088/0034-4885/68/10/R02} {\bibfield  {journal} {\bibinfo  {journal} {Rep. Prog. Phys.}\ }\textbf {\bibinfo {volume} {68}},\ \bibinfo {pages} {2337} (\bibinfo {year} {2005})}\BibitemShut {NoStop}%
\bibitem [{\citenamefont {v.~L{\"o}hneysen}(2011)}]{lohneysen11}%
  \BibitemOpen
  \bibfield  {author} {\bibinfo {author} {\bibfnamefont {H.}~\bibnamefont {v.~L{\"o}hneysen}},\ }\bibfield  {title} {\bibinfo {title} {Electron-electron interactions and the metal-insulator transition in heavily doped silicon},\ }\href {https://doi.org/https://doi.org/10.1002/andp.201100034} {\bibfield  {journal} {\bibinfo  {journal} {Ann. Phys. (Berlin)}\ }\textbf {\bibinfo {volume} {523}},\ \bibinfo {pages} {599} (\bibinfo {year} {2011})}\BibitemShut {NoStop}%
\bibitem [{\citenamefont {Dobrosavljevi{\'c}}\ \emph {et~al.}(2012)\citenamefont {Dobrosavljevi{\'c}}, \citenamefont {Trivedi},\ and\ \citenamefont {Valles}}]{vlad12}%
  \BibitemOpen
  \bibfield  {author} {\bibinfo {author} {\bibfnamefont {V.}~\bibnamefont {Dobrosavljevi{\'c}}}, \bibinfo {author} {\bibfnamefont {N.}~\bibnamefont {Trivedi}},\ and\ \bibinfo {author} {\bibfnamefont {J.}~\bibnamefont {Valles}},\ }\href@noop {} {\emph {\bibinfo {title} {Conductor Insulator Quantum Phase Transitions}}}\ (\bibinfo  {publisher} {OUP Oxford},\ \bibinfo {year} {2012})\BibitemShut {NoStop}%
\bibitem [{\citenamefont {Lakner}\ and\ \citenamefont {L\"ohneysen}(1989)}]{lakner89}%
  \BibitemOpen
  \bibfield  {author} {\bibinfo {author} {\bibfnamefont {M.}~\bibnamefont {Lakner}}\ and\ \bibinfo {author} {\bibfnamefont {H.~v.}\ \bibnamefont {L\"ohneysen}},\ }\bibfield  {title} {\bibinfo {title} {Localized versus itinerant electrons at the metal-insulator transition in si:p},\ }\href {https://doi.org/10.1103/PhysRevLett.63.648} {\bibfield  {journal} {\bibinfo  {journal} {Phys. Rev. Lett.}\ }\textbf {\bibinfo {volume} {63}},\ \bibinfo {pages} {648 } (\bibinfo {year} {1989})}\BibitemShut {NoStop}%
\bibitem [{\citenamefont {Paalanen}\ \emph {et~al.}(1988)\citenamefont {Paalanen}, \citenamefont {Graebner}, \citenamefont {Bhatt},\ and\ \citenamefont {Sachdev}}]{paalanen88}%
  \BibitemOpen
  \bibfield  {author} {\bibinfo {author} {\bibfnamefont {M.~A.}\ \bibnamefont {Paalanen}}, \bibinfo {author} {\bibfnamefont {J.~E.}\ \bibnamefont {Graebner}}, \bibinfo {author} {\bibfnamefont {R.~N.}\ \bibnamefont {Bhatt}},\ and\ \bibinfo {author} {\bibfnamefont {S.}~\bibnamefont {Sachdev}},\ }\bibfield  {title} {\bibinfo {title} {Thermodynamic behavior near a metal-insulator transition},\ }\href {https://doi.org/10.1103/PhysRevLett.61.597} {\bibfield  {journal} {\bibinfo  {journal} {Phys. Rev. Lett.}\ }\textbf {\bibinfo {volume} {61}},\ \bibinfo {pages} {597} (\bibinfo {year} {1988})}\BibitemShut {NoStop}%
\bibitem [{\citenamefont {Roy}\ and\ \citenamefont {Sarachik}(1988)}]{roy88}%
  \BibitemOpen
  \bibfield  {author} {\bibinfo {author} {\bibfnamefont {A.}~\bibnamefont {Roy}}\ and\ \bibinfo {author} {\bibfnamefont {M.~P.}\ \bibnamefont {Sarachik}},\ }\bibfield  {title} {\bibinfo {title} {Susceptibility of si:p across the metal-insulator transition. ii. evidence for local moments in the metallic phase},\ }\href {https://doi.org/10.1103/PhysRevB.37.5531} {\bibfield  {journal} {\bibinfo  {journal} {Phys. Rev. B}\ }\textbf {\bibinfo {volume} {37}},\ \bibinfo {pages} {5531} (\bibinfo {year} {1988})}\BibitemShut {NoStop}%
\bibitem [{\citenamefont {Milovanovi{\'c}}\ \emph {et~al.}(1989)\citenamefont {Milovanovi{\'c}}, \citenamefont {Sachdev},\ and\ \citenamefont {Bhatt}}]{milovanovic89}%
  \BibitemOpen
  \bibfield  {author} {\bibinfo {author} {\bibfnamefont {M.}~\bibnamefont {Milovanovi{\'c}}}, \bibinfo {author} {\bibfnamefont {S.}~\bibnamefont {Sachdev}},\ and\ \bibinfo {author} {\bibfnamefont {R.~N.}\ \bibnamefont {Bhatt}},\ }\bibfield  {title} {\bibinfo {title} {Effective-field theory of local-moment formation in disordered metals},\ }\href {https://doi.org/10.1103/PhysRevLett.63.82} {\bibfield  {journal} {\bibinfo  {journal} {Phys. Rev. Lett.}\ }\textbf {\bibinfo {volume} {63}},\ \bibinfo {pages} {82 } (\bibinfo {year} {1989})}\BibitemShut {NoStop}%
\bibitem [{\citenamefont {Bhatt}\ and\ \citenamefont {Fisher}(1992)}]{bhatt92}%
  \BibitemOpen
  \bibfield  {author} {\bibinfo {author} {\bibfnamefont {R.~N.}\ \bibnamefont {Bhatt}}\ and\ \bibinfo {author} {\bibfnamefont {D.~S.}\ \bibnamefont {Fisher}},\ }\bibfield  {title} {\bibinfo {title} {Absence of spin diffusion in most random lattices},\ }\href {https://doi.org/10.1103/PhysRevLett.68.3072} {\bibfield  {journal} {\bibinfo  {journal} {Phys. Rev. Lett.}\ }\textbf {\bibinfo {volume} {68}},\ \bibinfo {pages} {3072 } (\bibinfo {year} {1992})}\BibitemShut {NoStop}%
\bibitem [{\citenamefont {Byczuk}\ \emph {et~al.}(2005)\citenamefont {Byczuk}, \citenamefont {Hofstetter},\ and\ \citenamefont {Vollhardt}}]{byczuk05}%
  \BibitemOpen
  \bibfield  {author} {\bibinfo {author} {\bibfnamefont {K.}~\bibnamefont {Byczuk}}, \bibinfo {author} {\bibfnamefont {W.}~\bibnamefont {Hofstetter}},\ and\ \bibinfo {author} {\bibfnamefont {D.}~\bibnamefont {Vollhardt}},\ }\bibfield  {title} {\bibinfo {title} {Mott-hubbard transition versus anderson localization in correlated electron systems with disorder},\ }\href {https://doi.org/10.1103/PhysRevLett.94.056404} {\bibfield  {journal} {\bibinfo  {journal} {Phys. Rev. Lett.}\ }\textbf {\bibinfo {volume} {94}},\ \bibinfo {pages} {056404} (\bibinfo {year} {2005})}\BibitemShut {NoStop}%
\bibitem [{\citenamefont {Aguiar}\ \emph {et~al.}(2009)\citenamefont {Aguiar}, \citenamefont {Dobrosavljevi{\'c}}, \citenamefont {Abrahams},\ and\ \citenamefont {Kotliar}}]{carol09}%
  \BibitemOpen
  \bibfield  {author} {\bibinfo {author} {\bibfnamefont {M.~C.~O.}\ \bibnamefont {Aguiar}}, \bibinfo {author} {\bibfnamefont {V.}~\bibnamefont {Dobrosavljevi{\'c}}}, \bibinfo {author} {\bibfnamefont {E.}~\bibnamefont {Abrahams}},\ and\ \bibinfo {author} {\bibfnamefont {G.}~\bibnamefont {Kotliar}},\ }\bibfield  {title} {\bibinfo {title} {Critical behavior at the mott-anderson transition: A typical-medium theory perspective},\ }\href {https://doi.org/10.1103/PhysRevLett.102.156402} {\bibfield  {journal} {\bibinfo  {journal} {Phys. Rev. Lett.}\ }\textbf {\bibinfo {volume} {102}},\ \bibinfo {pages} {156402} (\bibinfo {year} {2009})}\BibitemShut {NoStop}%
\bibitem [{\citenamefont {Andrade}\ \emph {et~al.}(2009)\citenamefont {Andrade}, \citenamefont {Miranda},\ and\ \citenamefont {Dobrosavljevi{\'c}}}]{amd09}%
  \BibitemOpen
  \bibfield  {author} {\bibinfo {author} {\bibfnamefont {E.~C.}\ \bibnamefont {Andrade}}, \bibinfo {author} {\bibfnamefont {E.}~\bibnamefont {Miranda}},\ and\ \bibinfo {author} {\bibfnamefont {V.}~\bibnamefont {Dobrosavljevi{\'c}}},\ }\bibfield  {title} {\bibinfo {title} {Electronic griffiths phase of the $d = 2$ mott transition},\ }\href@noop {} {\bibfield  {journal} {\bibinfo  {journal} {Phys. Rev. Lett.}\ }\textbf {\bibinfo {volume} {102}},\ \bibinfo {pages} {206403} (\bibinfo {year} {2009})}\BibitemShut {NoStop}%
\bibitem [{\citenamefont {Abanin}\ \emph {et~al.}(2019)\citenamefont {Abanin}, \citenamefont {Altman}, \citenamefont {Bloch},\ and\ \citenamefont {Serbyn}}]{abanin19}%
  \BibitemOpen
  \bibfield  {author} {\bibinfo {author} {\bibfnamefont {D.~A.}\ \bibnamefont {Abanin}}, \bibinfo {author} {\bibfnamefont {E.}~\bibnamefont {Altman}}, \bibinfo {author} {\bibfnamefont {I.}~\bibnamefont {Bloch}},\ and\ \bibinfo {author} {\bibfnamefont {M.}~\bibnamefont {Serbyn}},\ }\bibfield  {title} {\bibinfo {title} {Colloquium: Many-body localization, thermalization, and entanglement},\ }\href {https://doi.org/10.1103/RevModPhys.91.021001} {\bibfield  {journal} {\bibinfo  {journal} {Rev. Mod. Phys.}\ }\textbf {\bibinfo {volume} {91}},\ \bibinfo {pages} {021001} (\bibinfo {year} {2019})}\BibitemShut {NoStop}%
\bibitem [{\citenamefont {Bhatt}\ and\ \citenamefont {Lee}(1982)}]{bhatt_lee82}%
  \BibitemOpen
  \bibfield  {author} {\bibinfo {author} {\bibfnamefont {R.~N.}\ \bibnamefont {Bhatt}}\ and\ \bibinfo {author} {\bibfnamefont {P.~A.}\ \bibnamefont {Lee}},\ }\bibfield  {title} {\bibinfo {title} {Scaling studies of highly disordered spin-\textonehalf{} antiferromagnetic systems},\ }\href {https://doi.org/10.1103/PhysRevLett.48.344} {\bibfield  {journal} {\bibinfo  {journal} {Phys. Rev. Lett.}\ }\textbf {\bibinfo {volume} {48}},\ \bibinfo {pages} {344} (\bibinfo {year} {1982})}\BibitemShut {NoStop}%
\bibitem [{\citenamefont {Dobrosavljevi{\'c}}\ \emph {et~al.}(1992)\citenamefont {Dobrosavljevi{\'c}}, \citenamefont {Kirkpatrick},\ and\ \citenamefont {Kotliar}}]{vlad92a}%
  \BibitemOpen
  \bibfield  {author} {\bibinfo {author} {\bibfnamefont {V.}~\bibnamefont {Dobrosavljevi{\'c}}}, \bibinfo {author} {\bibfnamefont {T.~R.}\ \bibnamefont {Kirkpatrick}},\ and\ \bibinfo {author} {\bibfnamefont {B.~G.}\ \bibnamefont {Kotliar}},\ }\bibfield  {title} {\bibinfo {title} {Kondo effect in disordered systems},\ }\href {https://doi.org/10.1103/PhysRevLett.69.1113} {\bibfield  {journal} {\bibinfo  {journal} {Phys. Rev. Lett.}\ }\textbf {\bibinfo {volume} {69}},\ \bibinfo {pages} {1113} (\bibinfo {year} {1992})}\BibitemShut {NoStop}%
\bibitem [{\citenamefont {Miranda}\ and\ \citenamefont {Dobrosavljevi{\'c}}(2001)}]{miranda01}%
  \BibitemOpen
  \bibfield  {author} {\bibinfo {author} {\bibfnamefont {E.}~\bibnamefont {Miranda}}\ and\ \bibinfo {author} {\bibfnamefont {V.}~\bibnamefont {Dobrosavljevi{\'c}}},\ }\bibfield  {title} {\bibinfo {title} {Localization-induced griffiths phase of disordered anderson lattices},\ }\href {https://doi.org/10.1103/PhysRevLett.86.264} {\bibfield  {journal} {\bibinfo  {journal} {Phys. Rev. Lett.}\ }\textbf {\bibinfo {volume} {86}},\ \bibinfo {pages} {264} (\bibinfo {year} {2001})}\BibitemShut {NoStop}%
\bibitem [{\citenamefont {Cornaglia}\ \emph {et~al.}(2006)\citenamefont {Cornaglia}, \citenamefont {Grempel},\ and\ \citenamefont {Balseiro}}]{cornaglia06}%
  \BibitemOpen
  \bibfield  {author} {\bibinfo {author} {\bibfnamefont {P.~S.}\ \bibnamefont {Cornaglia}}, \bibinfo {author} {\bibfnamefont {D.~R.}\ \bibnamefont {Grempel}},\ and\ \bibinfo {author} {\bibfnamefont {C.~A.}\ \bibnamefont {Balseiro}},\ }\bibfield  {title} {\bibinfo {title} {{Universal Distribution of Kondo Temperatures in Dirty Metals}},\ }\href {https://doi.org/10.1103/PhysRevLett.96.117209} {\bibfield  {journal} {\bibinfo  {journal} {Phys. Rev. Lett.}\ }\textbf {\bibinfo {volume} {96}},\ \bibinfo {pages} {117209} (\bibinfo {year} {2006})}\BibitemShut {NoStop}%
\bibitem [{\citenamefont {Kaul}\ and\ \citenamefont {Vojta}(2007)}]{kaul07}%
  \BibitemOpen
  \bibfield  {author} {\bibinfo {author} {\bibfnamefont {R.~K.}\ \bibnamefont {Kaul}}\ and\ \bibinfo {author} {\bibfnamefont {M.}~\bibnamefont {Vojta}},\ }\bibfield  {title} {\bibinfo {title} {{Strongly inhomogeneous phases and non-Fermi-liquid behavior in randomly depleted Kondo lattices}},\ }\href {https://doi.org/10.1103/PhysRevB.75.132407} {\bibfield  {journal} {\bibinfo  {journal} {Phys. Rev. B}\ }\textbf {\bibinfo {volume} {75}},\ \bibinfo {pages} {132407} (\bibinfo {year} {2007})}\BibitemShut {NoStop}%
\bibitem [{\citenamefont {Kettemann}\ \emph {et~al.}(2009)\citenamefont {Kettemann}, \citenamefont {Mucciolo},\ and\ \citenamefont {Varga}}]{kettemann09}%
  \BibitemOpen
  \bibfield  {author} {\bibinfo {author} {\bibfnamefont {S.}~\bibnamefont {Kettemann}}, \bibinfo {author} {\bibfnamefont {E.~R.}\ \bibnamefont {Mucciolo}},\ and\ \bibinfo {author} {\bibfnamefont {I.}~\bibnamefont {Varga}},\ }\bibfield  {title} {\bibinfo {title} {{Critical Metal Phase at the Anderson Metal-Insulator Transition with Kondo Impurities}},\ }\href {https://doi.org/10.1103/PhysRevLett.103.126401} {\bibfield  {journal} {\bibinfo  {journal} {Phys. Rev. Lett.}\ }\textbf {\bibinfo {volume} {103}},\ \bibinfo {pages} {126401} (\bibinfo {year} {2009})}\BibitemShut {NoStop}%
\bibitem [{\citenamefont {Kettemann}\ \emph {et~al.}(2012)\citenamefont {Kettemann}, \citenamefont {Mucciolo}, \citenamefont {Varga},\ and\ \citenamefont {Slevin}}]{kettemann12}%
  \BibitemOpen
  \bibfield  {author} {\bibinfo {author} {\bibfnamefont {S.}~\bibnamefont {Kettemann}}, \bibinfo {author} {\bibfnamefont {E.~R.}\ \bibnamefont {Mucciolo}}, \bibinfo {author} {\bibfnamefont {I.}~\bibnamefont {Varga}},\ and\ \bibinfo {author} {\bibfnamefont {K.}~\bibnamefont {Slevin}},\ }\bibfield  {title} {\bibinfo {title} {Kondo-anderson transitions},\ }\href {https://doi.org/10.1103/PhysRevB.85.115112} {\bibfield  {journal} {\bibinfo  {journal} {Phys. Rev. B}\ }\textbf {\bibinfo {volume} {85}},\ \bibinfo {pages} {115112} (\bibinfo {year} {2012})}\BibitemShut {NoStop}%
\bibitem [{\citenamefont {Miranda}\ \emph {et~al.}(2014)\citenamefont {Miranda}, \citenamefont {Dias~da Silva},\ and\ \citenamefont {Lewenkopf}}]{miranda14}%
  \BibitemOpen
  \bibfield  {author} {\bibinfo {author} {\bibfnamefont {V.~G.}\ \bibnamefont {Miranda}}, \bibinfo {author} {\bibfnamefont {L.~G. G.~V.}\ \bibnamefont {Dias~da Silva}},\ and\ \bibinfo {author} {\bibfnamefont {C.~H.}\ \bibnamefont {Lewenkopf}},\ }\bibfield  {title} {\bibinfo {title} {Disorder-mediated kondo effect in graphene},\ }\href {https://doi.org/10.1103/PhysRevB.90.201101} {\bibfield  {journal} {\bibinfo  {journal} {Phys. Rev. B}\ }\textbf {\bibinfo {volume} {90}},\ \bibinfo {pages} {201101} (\bibinfo {year} {2014})}\BibitemShut {NoStop}%
\bibitem [{\citenamefont {Andrade}\ \emph {et~al.}(2015)\citenamefont {Andrade}, \citenamefont {Jagannathan}, \citenamefont {Miranda}, \citenamefont {Vojta},\ and\ \citenamefont {Dobrosavljevi{\'c}}}]{andrade15}%
  \BibitemOpen
  \bibfield  {author} {\bibinfo {author} {\bibfnamefont {E.~C.}\ \bibnamefont {Andrade}}, \bibinfo {author} {\bibfnamefont {A.}~\bibnamefont {Jagannathan}}, \bibinfo {author} {\bibfnamefont {E.}~\bibnamefont {Miranda}}, \bibinfo {author} {\bibfnamefont {M.}~\bibnamefont {Vojta}},\ and\ \bibinfo {author} {\bibfnamefont {V.}~\bibnamefont {Dobrosavljevi{\'c}}},\ }\bibfield  {title} {\bibinfo {title} {Non-fermi-liquid behavior in metallic quasicrystals with local magnetic moments},\ }\href {https://doi.org/10.1103/PhysRevLett.115.036403} {\bibfield  {journal} {\bibinfo  {journal} {Phys. Rev. Lett.}\ }\textbf {\bibinfo {volume} {115}},\ \bibinfo {pages} {036403} (\bibinfo {year} {2015})}\BibitemShut {NoStop}%
\bibitem [{\citenamefont {Andres}\ \emph {et~al.}(1981)\citenamefont {Andres}, \citenamefont {Bhatt}, \citenamefont {Goalwin}, \citenamefont {Rice},\ and\ \citenamefont {Walstedt}}]{andres81}%
  \BibitemOpen
  \bibfield  {author} {\bibinfo {author} {\bibfnamefont {K.}~\bibnamefont {Andres}}, \bibinfo {author} {\bibfnamefont {R.~N.}\ \bibnamefont {Bhatt}}, \bibinfo {author} {\bibfnamefont {P.}~\bibnamefont {Goalwin}}, \bibinfo {author} {\bibfnamefont {T.~M.}\ \bibnamefont {Rice}},\ and\ \bibinfo {author} {\bibfnamefont {R.~E.}\ \bibnamefont {Walstedt}},\ }\bibfield  {title} {\bibinfo {title} {Low-temperature magnetic susceptibility of si: P in the nonmetallic region},\ }\href {https://doi.org/10.1103/PhysRevB.24.244} {\bibfield  {journal} {\bibinfo  {journal} {Phys. Rev. B}\ }\textbf {\bibinfo {volume} {24}},\ \bibinfo {pages} {244} (\bibinfo {year} {1981})}\BibitemShut {NoStop}%
\bibitem [{\citenamefont {Herring}\ and\ \citenamefont {Flicker}(1964)}]{herring64}%
  \BibitemOpen
  \bibfield  {author} {\bibinfo {author} {\bibfnamefont {C.}~\bibnamefont {Herring}}\ and\ \bibinfo {author} {\bibfnamefont {M.}~\bibnamefont {Flicker}},\ }\bibfield  {title} {\bibinfo {title} {Asymptotic exchange coupling of two hydrogen atoms},\ }\href {https://doi.org/10.1103/PhysRev.134.A362} {\bibfield  {journal} {\bibinfo  {journal} {Phys. Rev.}\ }\textbf {\bibinfo {volume} {134}},\ \bibinfo {pages} {A362} (\bibinfo {year} {1964})}\BibitemShut {NoStop}%
\bibitem [{\citenamefont {Koiller}\ \emph {et~al.}(2002)\citenamefont {Koiller}, \citenamefont {Hu},\ and\ \citenamefont {Das~Sarma}}]{koiller02}%
  \BibitemOpen
  \bibfield  {author} {\bibinfo {author} {\bibfnamefont {B.}~\bibnamefont {Koiller}}, \bibinfo {author} {\bibfnamefont {X.}~\bibnamefont {Hu}},\ and\ \bibinfo {author} {\bibfnamefont {S.}~\bibnamefont {Das~Sarma}},\ }\bibfield  {title} {\bibinfo {title} {Strain effects on silicon donor exchange: Quantum computer architecture considerations},\ }\href {https://doi.org/10.1103/PhysRevB.66.115201} {\bibfield  {journal} {\bibinfo  {journal} {Phys. Rev. B}\ }\textbf {\bibinfo {volume} {66}},\ \bibinfo {pages} {115201} (\bibinfo {year} {2002})}\BibitemShut {NoStop}%
\bibitem [{\citenamefont {Oganesyan}\ and\ \citenamefont {Huse}(2007)}]{oganesyan07}%
  \BibitemOpen
  \bibfield  {author} {\bibinfo {author} {\bibfnamefont {V.}~\bibnamefont {Oganesyan}}\ and\ \bibinfo {author} {\bibfnamefont {D.~A.}\ \bibnamefont {Huse}},\ }\bibfield  {title} {\bibinfo {title} {{Localization of interacting fermions at high temperature}},\ }\href {https://doi.org/10.1103/PhysRevB.75.155111} {\bibfield  {journal} {\bibinfo  {journal} {Phys. Rev. B}\ }\textbf {\bibinfo {volume} {75}},\ \bibinfo {pages} {155111} (\bibinfo {year} {2007})}\BibitemShut {NoStop}%
\bibitem [{\citenamefont {Atas}\ \emph {et~al.}(2013)\citenamefont {Atas}, \citenamefont {Bogomolny}, \citenamefont {Giraud},\ and\ \citenamefont {Roux}}]{atas13}%
  \BibitemOpen
  \bibfield  {author} {\bibinfo {author} {\bibfnamefont {Y.~Y.}\ \bibnamefont {Atas}}, \bibinfo {author} {\bibfnamefont {E.}~\bibnamefont {Bogomolny}}, \bibinfo {author} {\bibfnamefont {O.}~\bibnamefont {Giraud}},\ and\ \bibinfo {author} {\bibfnamefont {G.}~\bibnamefont {Roux}},\ }\bibfield  {title} {\bibinfo {title} {{Distribution of the Ratio of Consecutive Level Spacings in Random Matrix Ensembles}},\ }\href {https://doi.org/10.1103/PhysRevLett.110.084101} {\bibfield  {journal} {\bibinfo  {journal} {Phys. Rev. Lett.}\ }\textbf {\bibinfo {volume} {110}},\ \bibinfo {pages} {084101} (\bibinfo {year} {2013})}\BibitemShut {NoStop}%
\bibitem [{\citenamefont {Root}\ \emph {et~al.}(1988)\citenamefont {Root}, \citenamefont {Bauer},\ and\ \citenamefont {Skinner}}]{root88}%
  \BibitemOpen
  \bibfield  {author} {\bibinfo {author} {\bibfnamefont {L.~J.}\ \bibnamefont {Root}}, \bibinfo {author} {\bibfnamefont {J.~D.}\ \bibnamefont {Bauer}},\ and\ \bibinfo {author} {\bibfnamefont {J.~L.}\ \bibnamefont {Skinner}},\ }\bibfield  {title} {\bibinfo {title} {New approach to localization: Quantum connectivity},\ }\href {https://doi.org/10.1103/PhysRevB.37.5518} {\bibfield  {journal} {\bibinfo  {journal} {Phys. Rev. B}\ }\textbf {\bibinfo {volume} {37}},\ \bibinfo {pages} {5518} (\bibinfo {year} {1988})}\BibitemShut {NoStop}%
\bibitem [{\citenamefont {Krich}\ and\ \citenamefont {Aspuru-Guzik}(2011)}]{krich11}%
  \BibitemOpen
  \bibfield  {author} {\bibinfo {author} {\bibfnamefont {J.~J.}\ \bibnamefont {Krich}}\ and\ \bibinfo {author} {\bibfnamefont {A.}~\bibnamefont {Aspuru-Guzik}},\ }\bibfield  {title} {\bibinfo {title} {Scaling and localization lengths of a topologically disordered system},\ }\href {https://doi.org/10.1103/PhysRevLett.106.156405} {\bibfield  {journal} {\bibinfo  {journal} {Phys. Rev. Lett.}\ }\textbf {\bibinfo {volume} {106}},\ \bibinfo {pages} {156405} (\bibinfo {year} {2011})}\BibitemShut {NoStop}%
\bibitem [{\citenamefont {Mott}(1990)}]{mott_mit}%
  \BibitemOpen
  \bibfield  {author} {\bibinfo {author} {\bibfnamefont {N.}~\bibnamefont {Mott}},\ }\href@noop {} {\emph {\bibinfo {title} {Metal-insulator Transitions}}},\ \bibinfo {edition} {2nd}\ ed.\ (\bibinfo  {publisher} {Taylor \& Francis},\ \bibinfo {year} {1990})\BibitemShut {NoStop}%
\bibitem [{\citenamefont {Slevin}\ and\ \citenamefont {Ohtsuki}(1999)}]{slevin99}%
  \BibitemOpen
  \bibfield  {author} {\bibinfo {author} {\bibfnamefont {K.}~\bibnamefont {Slevin}}\ and\ \bibinfo {author} {\bibfnamefont {T.}~\bibnamefont {Ohtsuki}},\ }\bibfield  {title} {\bibinfo {title} {Corrections to scaling at the anderson transition},\ }\href {https://doi.org/10.1103/PhysRevLett.82.382} {\bibfield  {journal} {\bibinfo  {journal} {Phys. Rev. Lett.}\ }\textbf {\bibinfo {volume} {82}},\ \bibinfo {pages} {382} (\bibinfo {year} {1999})}\BibitemShut {NoStop}%
\bibitem [{\citenamefont {Dobrosavljevi{\'c}}\ \emph {et~al.}(2003)\citenamefont {Dobrosavljevi{\'c}}, \citenamefont {Pastor},\ and\ \citenamefont {Nikoli{\'c}}}]{vlad03}%
  \BibitemOpen
  \bibfield  {author} {\bibinfo {author} {\bibfnamefont {V.}~\bibnamefont {Dobrosavljevi{\'c}}}, \bibinfo {author} {\bibfnamefont {A.~A.}\ \bibnamefont {Pastor}},\ and\ \bibinfo {author} {\bibfnamefont {B.~K.}\ \bibnamefont {Nikoli{\'c}}},\ }\bibfield  {title} {\bibinfo {title} {Typical medium theory of anderson localization: A local order parameter approach to strong-disorder effects},\ }\href {https://doi.org/10.1209/epl/i2003-00364-5} {\bibfield  {journal} {\bibinfo  {journal} {EPL}\ }\textbf {\bibinfo {volume} {62}},\ \bibinfo {pages} {76} (\bibinfo {year} {2003})}\BibitemShut {NoStop}%
\bibitem [{\citenamefont {Markos}(2006)}]{makros06}%
  \BibitemOpen
  \bibfield  {author} {\bibinfo {author} {\bibfnamefont {P.}~\bibnamefont {Markos}},\ }\bibfield  {title} {\bibinfo {title} {Numerical analysis of the {Anderson} localization},\ }\href@noop {} {\bibfield  {journal} {\bibinfo  {journal} {Acta Phys. Slov.}\ }\textbf {\bibinfo {volume} {56}},\ \bibinfo {pages} {561} (\bibinfo {year} {2006})}\BibitemShut {NoStop}%
\bibitem [{\citenamefont {Rodriguez}\ \emph {et~al.}(2010)\citenamefont {Rodriguez}, \citenamefont {Vasquez}, \citenamefont {Slevin},\ and\ \citenamefont {R\"omer}}]{rodriguez10}%
  \BibitemOpen
  \bibfield  {author} {\bibinfo {author} {\bibfnamefont {A.}~\bibnamefont {Rodriguez}}, \bibinfo {author} {\bibfnamefont {L.~J.}\ \bibnamefont {Vasquez}}, \bibinfo {author} {\bibfnamefont {K.}~\bibnamefont {Slevin}},\ and\ \bibinfo {author} {\bibfnamefont {R.~A.}\ \bibnamefont {R\"omer}},\ }\bibfield  {title} {\bibinfo {title} {Critical parameters from a generalized multifractal analysis at the anderson transition},\ }\href {https://doi.org/10.1103/PhysRevLett.105.046403} {\bibfield  {journal} {\bibinfo  {journal} {Phys. Rev. Lett.}\ }\textbf {\bibinfo {volume} {105}},\ \bibinfo {pages} {046403} (\bibinfo {year} {2010})}\BibitemShut {NoStop}%
\bibitem [{\citenamefont {Denteneer}\ \emph {et~al.}(1999)\citenamefont {Denteneer}, \citenamefont {Scalettar},\ and\ \citenamefont {Trivedi}}]{denteneer99}%
  \BibitemOpen
  \bibfield  {author} {\bibinfo {author} {\bibfnamefont {P.~J.~H.}\ \bibnamefont {Denteneer}}, \bibinfo {author} {\bibfnamefont {R.~T.}\ \bibnamefont {Scalettar}},\ and\ \bibinfo {author} {\bibfnamefont {N.}~\bibnamefont {Trivedi}},\ }\bibfield  {title} {\bibinfo {title} {Conducting phase in the two-dimensional disordered hubbard model},\ }\href {https://doi.org/10.1103/PhysRevLett.83.4610} {\bibfield  {journal} {\bibinfo  {journal} {Phys. Rev. Lett.}\ }\textbf {\bibinfo {volume} {83}},\ \bibinfo {pages} {4610} (\bibinfo {year} {1999})}\BibitemShut {NoStop}%
\bibitem [{\citenamefont {Heidarian}\ and\ \citenamefont {Trivedi}(2004)}]{heidarian04}%
  \BibitemOpen
  \bibfield  {author} {\bibinfo {author} {\bibfnamefont {D.}~\bibnamefont {Heidarian}}\ and\ \bibinfo {author} {\bibfnamefont {N.}~\bibnamefont {Trivedi}},\ }\bibfield  {title} {\bibinfo {title} {Inhomogeneous metallic phase in a disordered mott insulator in two dimensions},\ }\href {https://doi.org/10.1103/PhysRevLett.93.126401} {\bibfield  {journal} {\bibinfo  {journal} {Phys. Rev. Lett.}\ }\textbf {\bibinfo {volume} {93}},\ \bibinfo {pages} {126401} (\bibinfo {year} {2004})}\BibitemShut {NoStop}%
\bibitem [{\citenamefont {{Dobrosavljevi\ifmmode \acute{c}\else {\'c}\fi{}}}\ and\ \citenamefont {Kotliar}(1997)}]{statdmft}%
  \BibitemOpen
  \bibfield  {author} {\bibinfo {author} {\bibfnamefont {V.}~\bibnamefont {{Dobrosavljevi\ifmmode \acute{c}\else {\'c}\fi{}}}}\ and\ \bibinfo {author} {\bibfnamefont {G.}~\bibnamefont {Kotliar}},\ }\bibfield  {title} {\bibinfo {title} {{Mean Field Theory of the Mott-Anderson Transition}},\ }\href {https://doi.org/10.1103/PhysRevLett.78.3943} {\bibfield  {journal} {\bibinfo  {journal} {Phys. Rev. Lett.}\ }\textbf {\bibinfo {volume} {78}},\ \bibinfo {pages} {3943} (\bibinfo {year} {1997})}\BibitemShut {NoStop}%
\bibitem [{\citenamefont {Ong}\ and\ \citenamefont {Jones}(2011)}]{ong11}%
  \BibitemOpen
  \bibfield  {author} {\bibinfo {author} {\bibfnamefont {T.~T.}\ \bibnamefont {Ong}}\ and\ \bibinfo {author} {\bibfnamefont {B.}~\bibnamefont {Jones}},\ }\bibfield  {title} {\bibinfo {title} {Generalized schrieffer-wolff transformation of the two-impurity kondo model},\ }\href@noop {} {\bibfield  {journal} {\bibinfo  {journal} {Europhys. Lett.}\ }\textbf {\bibinfo {volume} {93}},\ \bibinfo {pages} {57004} (\bibinfo {year} {2011})}\BibitemShut {NoStop}%
\bibitem [{\citenamefont {Thomas}\ \emph {et~al.}(1981)\citenamefont {Thomas}, \citenamefont {Capizzi}, \citenamefont {DeRosa}, \citenamefont {Bhatt},\ and\ \citenamefont {Rice}}]{thomas81}%
  \BibitemOpen
  \bibfield  {author} {\bibinfo {author} {\bibfnamefont {G.~A.}\ \bibnamefont {Thomas}}, \bibinfo {author} {\bibfnamefont {M.}~\bibnamefont {Capizzi}}, \bibinfo {author} {\bibfnamefont {F.}~\bibnamefont {DeRosa}}, \bibinfo {author} {\bibfnamefont {R.~N.}\ \bibnamefont {Bhatt}},\ and\ \bibinfo {author} {\bibfnamefont {T.~M.}\ \bibnamefont {Rice}},\ }\bibfield  {title} {\bibinfo {title} {Optical study of interacting donors in semiconductors},\ }\href {https://doi.org/10.1103/PhysRevB.23.5472} {\bibfield  {journal} {\bibinfo  {journal} {Phys. Rev. B}\ }\textbf {\bibinfo {volume} {23}},\ \bibinfo {pages} {5472} (\bibinfo {year} {1981})}\BibitemShut {NoStop}%
\bibitem [{\citenamefont {Jones}\ \emph {et~al.}(1989)\citenamefont {Jones}, \citenamefont {Kotliar},\ and\ \citenamefont {Millis}}]{jones89}%
  \BibitemOpen
  \bibfield  {author} {\bibinfo {author} {\bibfnamefont {B.~A.}\ \bibnamefont {Jones}}, \bibinfo {author} {\bibfnamefont {B.~G.}\ \bibnamefont {Kotliar}},\ and\ \bibinfo {author} {\bibfnamefont {A.~J.}\ \bibnamefont {Millis}},\ }\bibfield  {title} {\bibinfo {title} {Mean-field analysis of two antiferromagnetically coupled anderson impurities},\ }\href {https://doi.org/10.1103/PhysRevB.39.3415} {\bibfield  {journal} {\bibinfo  {journal} {Phys. Rev. B}\ }\textbf {\bibinfo {volume} {39}},\ \bibinfo {pages} {3415} (\bibinfo {year} {1989})}\BibitemShut {NoStop}%
\bibitem [{\citenamefont {Senthil}\ \emph {et~al.}(2004)\citenamefont {Senthil}, \citenamefont {Vojta},\ and\ \citenamefont {Sachdev}}]{senthil08}%
  \BibitemOpen
  \bibfield  {author} {\bibinfo {author} {\bibfnamefont {T.}~\bibnamefont {Senthil}}, \bibinfo {author} {\bibfnamefont {M.}~\bibnamefont {Vojta}},\ and\ \bibinfo {author} {\bibfnamefont {S.}~\bibnamefont {Sachdev}},\ }\bibfield  {title} {\bibinfo {title} {Weak magnetism and non-fermi liquids near heavy-fermion critical points},\ }\href {https://doi.org/10.1103/PhysRevB.69.035111} {\bibfield  {journal} {\bibinfo  {journal} {Phys. Rev. B}\ }\textbf {\bibinfo {volume} {69}},\ \bibinfo {pages} {035111} (\bibinfo {year} {2004})}\BibitemShut {NoStop}%
\bibitem [{\citenamefont {Hackl}\ and\ \citenamefont {Vojta}(2008)}]{hackl08}%
  \BibitemOpen
  \bibfield  {author} {\bibinfo {author} {\bibfnamefont {A.}~\bibnamefont {Hackl}}\ and\ \bibinfo {author} {\bibfnamefont {M.}~\bibnamefont {Vojta}},\ }\bibfield  {title} {\bibinfo {title} {Kondo volume collapse, kondo breakdown, and fermi surface transitions in heavy-fermion metals},\ }\href {https://doi.org/10.1103/PhysRevB.77.134439} {\bibfield  {journal} {\bibinfo  {journal} {Phys. Rev. B}\ }\textbf {\bibinfo {volume} {77}},\ \bibinfo {pages} {134439} (\bibinfo {year} {2008})}\BibitemShut {NoStop}%
\bibitem [{\citenamefont {Coleman}(2015)}]{coleman15}%
  \BibitemOpen
  \bibfield  {author} {\bibinfo {author} {\bibfnamefont {P.}~\bibnamefont {Coleman}},\ }\href@noop {} {\emph {\bibinfo {title} {Introduction to Many-Body Physics}}}\ (\bibinfo  {publisher} {Cambridge University Press},\ \bibinfo {year} {2015})\BibitemShut {NoStop}%
\bibitem [{sup()}]{suppl}%
  \BibitemOpen
  \href@noop {} {}\bibinfo {note} {See Supplemental Material at \url{https://link.aps.org/supplemental/10.1103/9fn5-mlwg} for additional details on the noninteracting problem, variational solution methodology, extended results, and discussion, which includes Refs.~ \cite{oganesyan07,atas13,hoyos04,clauset09,vlad92b}.}\BibitemShut {Stop}%
\bibitem [{\citenamefont {Hewson}(1993)}]{Hewson_kondo}%
  \BibitemOpen
  \bibfield  {author} {\bibinfo {author} {\bibfnamefont {A.~C.}\ \bibnamefont {Hewson}},\ }\href@noop {} {\emph {\bibinfo {title} {The Kondo Problem to Heavy Fermions}}},\ \bibinfo {edition} {1st}\ ed.\ (\bibinfo  {publisher} {Cambridge University Press},\ \bibinfo {address} {Cambridge},\ \bibinfo {year} {1993})\BibitemShut {NoStop}%
\bibitem [{\citenamefont {Dong}\ \emph {et~al.}(2021)\citenamefont {Dong}, \citenamefont {Huang},\ and\ \citenamefont {Yang}}]{dong21}%
  \BibitemOpen
  \bibfield  {author} {\bibinfo {author} {\bibfnamefont {J.-J.}\ \bibnamefont {Dong}}, \bibinfo {author} {\bibfnamefont {D.}~\bibnamefont {Huang}},\ and\ \bibinfo {author} {\bibfnamefont {Y.-f.}\ \bibnamefont {Yang}},\ }\bibfield  {title} {\bibinfo {title} {Mutual information, quantum phase transition, and phase coherence in kondo systems},\ }\href {https://doi.org/10.1103/PhysRevB.104.L081115} {\bibfield  {journal} {\bibinfo  {journal} {Phys. Rev. B}\ }\textbf {\bibinfo {volume} {104}},\ \bibinfo {pages} {L081115} (\bibinfo {year} {2021})}\BibitemShut {NoStop}%
\bibitem [{\citenamefont {Fye}(1994)}]{fye94}%
  \BibitemOpen
  \bibfield  {author} {\bibinfo {author} {\bibfnamefont {R.~M.}\ \bibnamefont {Fye}},\ }\bibfield  {title} {\bibinfo {title} {``anomalous fixed point behavior'' of two kondo impurities: A reexamination},\ }\href {https://doi.org/10.1103/PhysRevLett.72.916} {\bibfield  {journal} {\bibinfo  {journal} {Phys. Rev. Lett.}\ }\textbf {\bibinfo {volume} {72}},\ \bibinfo {pages} {916} (\bibinfo {year} {1994})}\BibitemShut {NoStop}%
\bibitem [{\citenamefont {Affleck}\ \emph {et~al.}(1995)\citenamefont {Affleck}, \citenamefont {Ludwig},\ and\ \citenamefont {Jones}}]{affleck95}%
  \BibitemOpen
  \bibfield  {author} {\bibinfo {author} {\bibfnamefont {I.}~\bibnamefont {Affleck}}, \bibinfo {author} {\bibfnamefont {A.~W.~W.}\ \bibnamefont {Ludwig}},\ and\ \bibinfo {author} {\bibfnamefont {B.~A.}\ \bibnamefont {Jones}},\ }\bibfield  {title} {\bibinfo {title} {Conformal-field-theory approach to the two-impurity kondo problem: Comparison with numerical renormalization-group results},\ }\href {https://doi.org/10.1103/PhysRevB.52.9528} {\bibfield  {journal} {\bibinfo  {journal} {Phys. Rev. B}\ }\textbf {\bibinfo {volume} {52}},\ \bibinfo {pages} {9528} (\bibinfo {year} {1995})}\BibitemShut {NoStop}%
\bibitem [{\citenamefont {Zhou}\ \emph {et~al.}(2009)\citenamefont {Zhou}, \citenamefont {Hoyos}, \citenamefont {Dobrosavljevi{\'c}},\ and\ \citenamefont {Miranda}}]{zhou09}%
  \BibitemOpen
  \bibfield  {author} {\bibinfo {author} {\bibfnamefont {S.}~\bibnamefont {Zhou}}, \bibinfo {author} {\bibfnamefont {J.~A.}\ \bibnamefont {Hoyos}}, \bibinfo {author} {\bibfnamefont {V.}~\bibnamefont {Dobrosavljevi{\'c}}},\ and\ \bibinfo {author} {\bibfnamefont {E.}~\bibnamefont {Miranda}},\ }\bibfield  {title} {\bibinfo {title} {Valence-bond theory of highly disordered quantum antiferromagnets},\ }\href {https://doi.org/10.1209/0295-5075/87/27003} {\bibfield  {journal} {\bibinfo  {journal} {Europhys. Lett.}\ }\textbf {\bibinfo {volume} {87}},\ \bibinfo {pages} {27003} (\bibinfo {year} {2009})}\BibitemShut {NoStop}%
\bibitem [{\citenamefont {Tanaskovi{\'c}}\ \emph {et~al.}(2004)\citenamefont {Tanaskovi{\'c}}, \citenamefont {Miranda},\ and\ \citenamefont {Dobrosavljevi{\'c}}}]{darko04}%
  \BibitemOpen
  \bibfield  {author} {\bibinfo {author} {\bibfnamefont {D.}~\bibnamefont {Tanaskovi{\'c}}}, \bibinfo {author} {\bibfnamefont {E.}~\bibnamefont {Miranda}},\ and\ \bibinfo {author} {\bibfnamefont {V.}~\bibnamefont {Dobrosavljevi{\'c}}},\ }\bibfield  {title} {\bibinfo {title} {Effective model of the electronic griffiths phase},\ }\href {https://doi.org/10.1103/PhysRevB.70.205108} {\bibfield  {journal} {\bibinfo  {journal} {Phys. Rev. B}\ }\textbf {\bibinfo {volume} {70}},\ \bibinfo {pages} {205108} (\bibinfo {year} {2004})}\BibitemShut {NoStop}%
\bibitem [{\citenamefont {Sachdev}(2023)}]{sachdev2023}%
  \BibitemOpen
  \bibfield  {author} {\bibinfo {author} {\bibfnamefont {S.}~\bibnamefont {Sachdev}},\ }\href {https://doi.org/10.1017/9781009212717} {\emph {\bibinfo {title} {Quantum Phases of Matter}}}\ (\bibinfo  {publisher} {Cambridge University Press},\ \bibinfo {year} {2023})\BibitemShut {NoStop}%
\bibitem [{\citenamefont {Kettemann}(2023)}]{kettemann23}%
  \BibitemOpen
  \bibfield  {author} {\bibinfo {author} {\bibfnamefont {S.}~\bibnamefont {Kettemann}},\ }\bibfield  {title} {\bibinfo {title} {Towards a comprehensive theory of metal-insulator transitions in doped semiconductors},\ }\href {https://doi.org/https://doi.org/10.1016/j.aop.2023.169306} {\bibfield  {journal} {\bibinfo  {journal} {Ann. Phys.}\ }\textbf {\bibinfo {volume} {456}},\ \bibinfo {pages} {169306} (\bibinfo {year} {2023})}\BibitemShut {NoStop}%
\bibitem [{\citenamefont {Javan~Mard}\ \emph {et~al.}(2015)\citenamefont {Javan~Mard}, \citenamefont {Andrade}, \citenamefont {Miranda},\ and\ \citenamefont {Dobrosavljevi\'c}}]{mard15}%
  \BibitemOpen
  \bibfield  {author} {\bibinfo {author} {\bibfnamefont {H.}~\bibnamefont {Javan~Mard}}, \bibinfo {author} {\bibfnamefont {E.~C.}\ \bibnamefont {Andrade}}, \bibinfo {author} {\bibfnamefont {E.}~\bibnamefont {Miranda}},\ and\ \bibinfo {author} {\bibfnamefont {V.}~\bibnamefont {Dobrosavljevi\'c}},\ }\bibfield  {title} {\bibinfo {title} {Non-gaussian spatial correlations dramatically weaken localization},\ }\href {https://doi.org/10.1103/PhysRevLett.114.056401} {\bibfield  {journal} {\bibinfo  {journal} {Phys. Rev. Lett.}\ }\textbf {\bibinfo {volume} {114}},\ \bibinfo {pages} {056401} (\bibinfo {year} {2015})}\BibitemShut {NoStop}%
\bibitem [{\citenamefont {Burdin}\ \emph {et~al.}(2000)\citenamefont {Burdin}, \citenamefont {Georges},\ and\ \citenamefont {Grempel}}]{burdin00}%
  \BibitemOpen
  \bibfield  {author} {\bibinfo {author} {\bibfnamefont {S.}~\bibnamefont {Burdin}}, \bibinfo {author} {\bibfnamefont {A.}~\bibnamefont {Georges}},\ and\ \bibinfo {author} {\bibfnamefont {D.~R.}\ \bibnamefont {Grempel}},\ }\bibfield  {title} {\bibinfo {title} {{Coherence Scale of the Kondo Lattice}},\ }\href {https://doi.org/10.1103/PhysRevLett.85.1048} {\bibfield  {journal} {\bibinfo  {journal} {Phys. Rev. Lett.}\ }\textbf {\bibinfo {volume} {85}},\ \bibinfo {pages} {1048} (\bibinfo {year} {2000})}\BibitemShut {NoStop}%
\bibitem [{\citenamefont {Oshikawa}(2000)}]{oshikawa00}%
  \BibitemOpen
  \bibfield  {author} {\bibinfo {author} {\bibfnamefont {M.}~\bibnamefont {Oshikawa}},\ }\bibfield  {title} {\bibinfo {title} {Topological approach to luttinger's theorem and the fermi surface of a kondo lattice},\ }\href {https://doi.org/10.1103/PhysRevLett.84.3370} {\bibfield  {journal} {\bibinfo  {journal} {Phys. Rev. Lett.}\ }\textbf {\bibinfo {volume} {84}},\ \bibinfo {pages} {3370} (\bibinfo {year} {2000})}\BibitemShut {NoStop}%
\bibitem [{\citenamefont {Senthil}\ \emph {et~al.}(2003)\citenamefont {Senthil}, \citenamefont {Sachdev},\ and\ \citenamefont {Vojta}}]{senthil03}%
  \BibitemOpen
  \bibfield  {author} {\bibinfo {author} {\bibfnamefont {T.}~\bibnamefont {Senthil}}, \bibinfo {author} {\bibfnamefont {S.}~\bibnamefont {Sachdev}},\ and\ \bibinfo {author} {\bibfnamefont {M.}~\bibnamefont {Vojta}},\ }\bibfield  {title} {\bibinfo {title} {Fractionalized fermi liquids},\ }\href {https://doi.org/10.1103/PhysRevLett.90.216403} {\bibfield  {journal} {\bibinfo  {journal} {Phys. Rev. Lett.}\ }\textbf {\bibinfo {volume} {90}},\ \bibinfo {pages} {216403} (\bibinfo {year} {2003})}\BibitemShut {NoStop}%
\bibitem [{\citenamefont {Lechermann}\ \emph {et~al.}(2007)\citenamefont {Lechermann}, \citenamefont {Georges}, \citenamefont {Kotliar},\ and\ \citenamefont {Parcollet}}]{lechermann07}%
  \BibitemOpen
  \bibfield  {author} {\bibinfo {author} {\bibfnamefont {F.}~\bibnamefont {Lechermann}}, \bibinfo {author} {\bibfnamefont {A.}~\bibnamefont {Georges}}, \bibinfo {author} {\bibfnamefont {G.}~\bibnamefont {Kotliar}},\ and\ \bibinfo {author} {\bibfnamefont {O.}~\bibnamefont {Parcollet}},\ }\bibfield  {title} {\bibinfo {title} {Rotationally invariant slave-boson formalism and momentum dependence of the quasiparticle weight},\ }\href {https://doi.org/10.1103/PhysRevB.76.155102} {\bibfield  {journal} {\bibinfo  {journal} {Phys. Rev. B}\ }\textbf {\bibinfo {volume} {76}},\ \bibinfo {pages} {155102} (\bibinfo {year} {2007})}\BibitemShut {NoStop}%
\bibitem [{\citenamefont {Lee}\ \emph {et~al.}(2016)\citenamefont {Lee}, \citenamefont {Florens},\ and\ \citenamefont {Dobrosavljevi\'c}}]{lee16}%
  \BibitemOpen
  \bibfield  {author} {\bibinfo {author} {\bibfnamefont {T.-H.}\ \bibnamefont {Lee}}, \bibinfo {author} {\bibfnamefont {S.}~\bibnamefont {Florens}},\ and\ \bibinfo {author} {\bibfnamefont {V.}~\bibnamefont {Dobrosavljevi\'c}},\ }\bibfield  {title} {\bibinfo {title} {Fate of spinons at the mott point},\ }\href {https://doi.org/10.1103/PhysRevLett.117.136601} {\bibfield  {journal} {\bibinfo  {journal} {Phys. Rev. Lett.}\ }\textbf {\bibinfo {volume} {117}},\ \bibinfo {pages} {136601} (\bibinfo {year} {2016})}\BibitemShut {NoStop}%
\bibitem [{\citenamefont {N\'ajera}\ \emph {et~al.}(2017)\citenamefont {N\'ajera}, \citenamefont {Civelli}, \citenamefont {Dobrosavljevi\'c},\ and\ \citenamefont {Rozenberg}}]{najera17}%
  \BibitemOpen
  \bibfield  {author} {\bibinfo {author} {\bibfnamefont {O.}~\bibnamefont {N\'ajera}}, \bibinfo {author} {\bibfnamefont {M.}~\bibnamefont {Civelli}}, \bibinfo {author} {\bibfnamefont {V.}~\bibnamefont {Dobrosavljevi\'c}},\ and\ \bibinfo {author} {\bibfnamefont {M.~J.}\ \bibnamefont {Rozenberg}},\ }\bibfield  {title} {\bibinfo {title} {Resolving the ${\mathrm{vo}}_{2}$ controversy: Mott mechanism dominates the insulator-to-metal transition},\ }\href {https://doi.org/10.1103/PhysRevB.95.035113} {\bibfield  {journal} {\bibinfo  {journal} {Phys. Rev. B}\ }\textbf {\bibinfo {volume} {95}},\ \bibinfo {pages} {035113} (\bibinfo {year} {2017})}\BibitemShut {NoStop}%
\bibitem [{\citenamefont {Drechsler}\ and\ \citenamefont {Vojta}(2025)}]{drechsler25}%
  \BibitemOpen
  \bibfield  {author} {\bibinfo {author} {\bibfnamefont {T.}~\bibnamefont {Drechsler}}\ and\ \bibinfo {author} {\bibfnamefont {M.}~\bibnamefont {Vojta}},\ }\bibfield  {title} {\bibinfo {title} {Emergent chiral metal near a kondo breakdown quantum phase transition},\ }\href {https://doi.org/10.1103/PhysRevLett.134.106503} {\bibfield  {journal} {\bibinfo  {journal} {Phys. Rev. Lett.}\ }\textbf {\bibinfo {volume} {134}},\ \bibinfo {pages} {106503} (\bibinfo {year} {2025})}\BibitemShut {NoStop}%
\bibitem [{\citenamefont {Rabelo}\ \emph {et~al.}(2026)\citenamefont {Rabelo}, \citenamefont {Almeida}, \citenamefont {Miranda}, \citenamefont {Dobrosavljevi{\'c}},\ and\ \citenamefont {Andrade}}]{Rabelo2026data}%
  \BibitemOpen
  \bibfield  {author} {\bibinfo {author} {\bibfnamefont {L.~G.}\ \bibnamefont {Rabelo}}, \bibinfo {author} {\bibfnamefont {I.~C.}\ \bibnamefont {Almeida}}, \bibinfo {author} {\bibfnamefont {E.}~\bibnamefont {Miranda}}, \bibinfo {author} {\bibfnamefont {V.}~\bibnamefont {Dobrosavljevi{\'c}}},\ and\ \bibinfo {author} {\bibfnamefont {E.~C.}\ \bibnamefont {Andrade}},\ }\bibfield  {title} {\bibinfo {title} {Numerical data from ``kondo screening and random-singlet formation in highly disordered systems''},\ }\href {https://doi.org/10.5281/zenodo.23069541} {10.5281/zenodo.23069541} (\bibinfo {year} {2026}),\ \bibinfo {note} {[Dataset], Zenodo}\BibitemShut {NoStop}%
\bibitem [{\citenamefont {Hoyos}\ and\ \citenamefont {Miranda}(2004)}]{hoyos04}%
  \BibitemOpen
  \bibfield  {author} {\bibinfo {author} {\bibfnamefont {J.~A.}\ \bibnamefont {Hoyos}}\ and\ \bibinfo {author} {\bibfnamefont {E.}~\bibnamefont {Miranda}},\ }\bibfield  {title} {\bibinfo {title} {Random antiferromagnetic $\mathrm{SU}(n)$ spin chains},\ }\href {https://doi.org/10.1103/PhysRevB.70.180401} {\bibfield  {journal} {\bibinfo  {journal} {Phys. Rev. B}\ }\textbf {\bibinfo {volume} {70}},\ \bibinfo {pages} {180401(R)} (\bibinfo {year} {2004})}\BibitemShut {NoStop}%
\bibitem [{\citenamefont {Clauset}\ \emph {et~al.}(2009)\citenamefont {Clauset}, \citenamefont {Shalizi},\ and\ \citenamefont {Newman}}]{clauset09}%
  \BibitemOpen
  \bibfield  {author} {\bibinfo {author} {\bibfnamefont {A.}~\bibnamefont {Clauset}}, \bibinfo {author} {\bibfnamefont {C.~R.}\ \bibnamefont {Shalizi}},\ and\ \bibinfo {author} {\bibfnamefont {M.~E.~J.}\ \bibnamefont {Newman}},\ }\bibfield  {title} {\bibinfo {title} {Power-law distributions in empirical data},\ }\href {https://doi.org/10.1137/070710111} {\bibfield  {journal} {\bibinfo  {journal} {SIAM Review}\ }\textbf {\bibinfo {volume} {51}},\ \bibinfo {pages} {661} (\bibinfo {year} {2009})}\BibitemShut {NoStop}%
\bibitem [{\citenamefont {Dobrosavljevi\ifmmode~\acute{c}\else \'{c}\fi{}}\ and\ \citenamefont {Kotliar}(1992)}]{vlad92b}%
  \BibitemOpen
  \bibfield  {author} {\bibinfo {author} {\bibfnamefont {V.}~\bibnamefont {Dobrosavljevi\ifmmode~\acute{c}\else \'{c}\fi{}}}\ and\ \bibinfo {author} {\bibfnamefont {G.}~\bibnamefont {Kotliar}},\ }\bibfield  {title} {\bibinfo {title} {Strong-coupling limit of the kondo problem},\ }\href {https://doi.org/10.1103/PhysRevB.46.5366} {\bibfield  {journal} {\bibinfo  {journal} {Phys. Rev. B}\ }\textbf {\bibinfo {volume} {46}},\ \bibinfo {pages} {5366} (\bibinfo {year} {1992})}\BibitemShut {NoStop}%
\end{thebibliography}
\end{document}

\title{Supplemental material for: \\Kondo screening and random-singlet formation in highly disordered systems }

\author{Lucas G.  Rabelo}
\affiliation{Instituto de F\'isica, Universidade de S\~ao Paulo, 05315-970, S\~ao Paulo, SP, Brazil}
\author{Igor C. Almeida}
\affiliation{Instituto de F\'isica de S\~ao Carlos, Universidade de S\~ao Paulo, S\~ao Carlos, SP, 13560-970, Brazil}
\author{Eduardo Miranda}
\affiliation{Gleb Wataghin Physics Institute, The University of Campinas, Rua S\'ergio Buarque de Holanda, 777, CEP 13083-859 Campinas, SP, Brazil}
\author{Vladimir Dobrosavljevi\'c}
\affiliation{Department of Physics and National High Magnetic Field Laboratory, Florida State University, Tallahassee, FL 32306}
\author{Eric C. Andrade}
\affiliation{Instituto de F\'isica, Universidade de S\~ao Paulo, 05315-970, S\~ao Paulo, SP, Brazil}

\date{\today}
\maketitle

\section{Further details of the model and of the large$-\mathcal{N}$ solution}

We present further details of the mean-field solution of the two-impurity Kondo problem.

\subsection{Clean case}

To illustrate the physical content of our solution to the two-impurity Kondo problem, we study two magnetic impurities interacting with a clean (homogeneous) conduction electron bath. For simplicity,  we work on a square lattice with nearest-neighbor hopping $t$ and assume local Kondo couplings, $K_{ij}^a=J_K^a \delta_{iA}\delta_{jA}$ and $K_{ij}^b=J_{K}^b\delta_{iB}\delta_{jB}$, where $J_K^{a}=8t_{aA}^2/U$ and $J_K^b=8t_{bB}^2/U$. The local assumption implies that impurity $a$ ($b$) couples only to the lattice site $\ket{A}$ ($\ket{B}$) located at $\bm{r}_A$ ($\bm{r}_B$).  Importantly, in the single-impurity regime --- where a single local moment $d$ couples locally to the origin site $\ket{o}$ and $|\Delta|=0$ --- Eq. A4 of the main text reduces to the well-known Nagaoka--Suhl one-loop equation \cite{Hewson_kondo}
\begin{equation}
    \sum_{\nu} \frac{\left|\braket{o|\nu}\right|^2}{E_\nu} \tanh\left(\frac{E_\nu}{2T_{K,d}^0} \right)=\frac{2}{J_K},
\end{equation}
where $T_{K,d}^0$ is the single-impurity Kondo temperature for the local moment $d$. 

\begin{figure}[t]
\begin{centering}
\includegraphics[width=1\columnwidth]{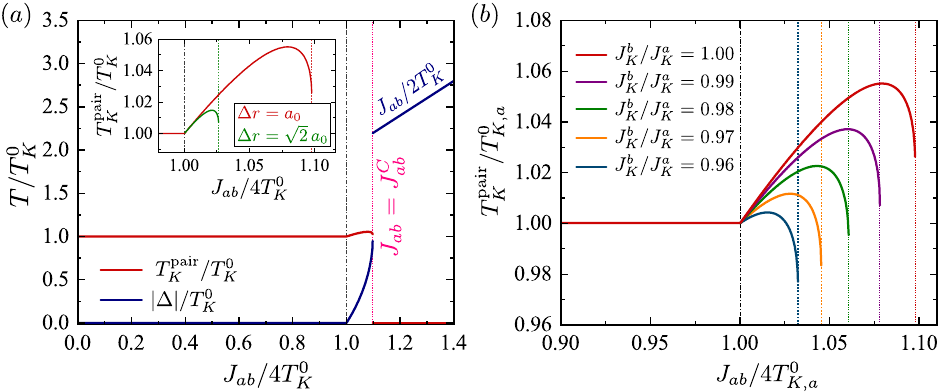}
\par\end{centering}
\caption{\label{fig:clean2} Solution of the two-impurity Kondo problem in the clean limit.  We model the conduction electrons as a square lattice with nearest neighbor hopping $t$ and with $\mu=-0.1t$.  (a) Phase diagram of the model with a fixed local Kondo coupling, $J_K/t=1$, as a function of the Heisenberg exchange $J_{ab}$, considering impurity $a$ $(b)$ connected locally to the lattice site at $\bm{r}_A$ ($\bm{r}_B$), with $\Delta r=|\bm{r}_A-\bm{r}_B|=a_0$. Here, $T_K^0=2.88\times 10^{-2}\,t$. Inset: Kondo temperature $T_K^\mathrm{pair}$ as a function of $J_{ab}/4T_K^0$ for two distinct values of $\Delta r$. (b) Kondo temperature of the model with distinct local Kondo couplings, $J_K^a \ne J_K^b$, and $\Delta r=a_0$, showing the rapid suppression of the coexistence region with increasing asymmetry $|J_K^a-J_K^b|$. Here, dotted lines indicate the critical Heisenberg coupling where $T_K^\mathrm{pair}$ abruptly drops to zero. The dashed-dotted line marks the condition $J_{ab}=4T_{K,a}^{0}$, where $T_{K,a}^0= 2.88\times 10^{-2}\,t$ is the Kondo temperature calculated at $J_{ab}=0$. }
\end{figure}

We begin by analyzing the two-impurity Kondo problem in the uniform case $J_K^a=J_K^b=J_K$.  Fig.~\ref{fig:clean2}(a) shows the resulting phase diagram at a fixed local Kondo coupling $J_K/t=1$ and impurity separation $\Delta r=|\bm{r}_A-\bm{r}_B|=a_0$, where $a_0$ is the lattice spacing for the periodic lattice discussed in this example.  For $J_{ab}=0$, we get a finite Kondo temperature $T_K^\mathrm{pair}=T_K^0$ and $|\Delta|=0$. This is the single-impurity solution,  where both impurities are individually quenched by the conduction electrons below the same Kondo temperature $T_K^0$.  As a function of $J_{ab}$, the Kondo temperature remains unaltered until $J_{ab}=4T_K^0\equiv J_{ab}^*$. Beyond this point, we find a narrow window, $J_{ab}\in [J_{ab}^*,J_{ab}^C]$, where both $T_K^\mathrm{pair}$ and $|\Delta|$ are non-vanishing.  Above a critical value $J_{ab}^C$,  the system enters the non-Kondo (random-singlet) state with $|\Delta|=J_{ab}/2$ and $T_K^\mathrm{pair}=0$.  In general,  this transition is discontinuous.  In the coexistence region, there is a competition between Kondo screening (scale $T_K^\mathrm{pair}$) and random-singlet formation (scale $|\Delta|$), and we observe a slight
initial increase in $T_K$.  This occurs because the Kondo resonance at the Fermi level (for $\Delta=0$) splits into two peaks centered at $\pm{\Delta}$. The Kondo temperature, defined as the half-width at half-maximum, thus increases. As we tune up $J_{ab}$, the DOS at the Fermi level is suppressed, and the Kondo effect disappears \cite{vlad92b}.  We also note that the coexistence region is drastically suppressed as a function of impurity separation [see inset of Fig.~\ref{fig:clean2}(a)].  

%For $J_{ab}>J_{ab}^C$, we recover the non-Kondo or random-singlet solution, $T_K^\mathrm{pair}=0$ and $|\Delta|=J_{ab}/2$,  where the local moments lock into a random-singlet state.  In our phase diagrams, $J_{ab}^C$ is consistent with the Varma-Jones critical point \citep{jones88,jones89}. For $\Delta r=a_\mathrm{lat}$ and several values of $J_K$, we find the relation $(J_{ab}^C/T_K^0)\approx 4.4$, roughly twice the Varma-Jones criterion. We attribute this discrepancy to their $T_K^0$ convention. 

We now investigate the clean case with different local Kondo couplings, $J_K^a\ne J_K^b$, and fix $J_K^a\ge J_K^b$ (hence $T_{K,a}^0 \ge T_{K,b}^0$). As shown in Fig.~\ref{fig:clean2}(b), for $J_{ab} \le 4T_{K,a}^0$, we obtain the single-impurity solution.  In this case, $|\Delta|=0$ and both moments are Kondo quenched independently, each below its own single-impurity Kondo temperature.  For $J_{ab}>4T_{K,a}^0$, we observe the emergence of a narrow coexistence region where $\Delta\ne0$ and $T_K^{\mathrm{pair}}>0$.  The larger single-impurity Kondo temperature sets the scale in this region,  whose width shrinks rapidly as the asymmetry $|J_K^a-J_K^b|$ increases. 

%In the coexistence region, Kondo screening competes with random-singlet formation. For $T_K^\mathrm{pair}>|\Delta|$, both impurities are quenched below the same Kondo temperature $T_K^\mathrm{pair}$, whereas for $T_K^\mathrm{pair}<|\Delta|$ a random-singlet state is favored. Above a critical value $J_{ab}^C$ [dotted lines in Fig.~\ref{fig:clean2}(b)], the system enters the non-Kondo (random-singlet) state with $|\Delta|=J_{ab}/2$ and $T_K^\mathrm{pair}=0$.

Taken together, these results show that the two-impurity phase diagram is dominated by $(i)$ a single-impurity regime with $\Delta=0$, where the local moments $a$ and $b$ are individually quenched by the conduction electrons below $T_{K,a}^0$ and $T_{K,b}^0$, respectively, and $(ii)$ a random-singlet regime, in which the local moments pair into a random-singlet state characterized by $|\Delta|=J_{ab}/2$ and $T_K^{\mathrm{pair}}=0$. The coexistence solution that captures the competition between Kondo screening and random-singlet formation, where both $\Delta$ and $T_K^\mathrm{pair}$ are finite, occurs only within a narrow parameter window and is rapidly suppressed as the impurity separation increases and/or the Kondo temperatures become asymmetric.

%$(c)$ Phase diagram for a fixed Heisenberg exchange, $J_{ab}/t=0.2$, as a function of the Kondo coupling $J_{ab}$ and $\Delta r=a$, and (d) Kondo temperature $T_K$ as a function of $J_{K}$ for two distinct distances $\Delta r$.
\subsection{Disordered case}

To solve the problem in the disordered case, we first determine the single-impurity Kondo temperature at every dopant site. For each of the $N_\mathrm{imp}$ impurities, we consider an independent Kondo problem in which the chosen site acts as a local moment while the remaining $N_\mathrm{imp}-1$ sites form the conduction bath. As a result, for every impurity choice, we generate a distinct conduction band Hamiltonian $\mathcal{H}_0$.  Because our simulations employ $N_\mathrm{imp}\gg 1$, the spectra of $\mathcal{H}_0$ with $N_\mathrm{imp}-1$ or $N_\mathrm{imp}-2$ sites are virtually identical, ensuring that the single- and two-impurity calculations can be compared on an equal footing. 

%I suggest a footnote here: this procedure is only ill-defined for the first pairs in the Kondo base; nevertheless, they are either Kondo screened with high Kondo temperatures or form RS pairs with significant gaps that are not important for the low-energy physics. 

\begin{figure}[t]
\begin{centering}
\includegraphics[width=1\columnwidth]{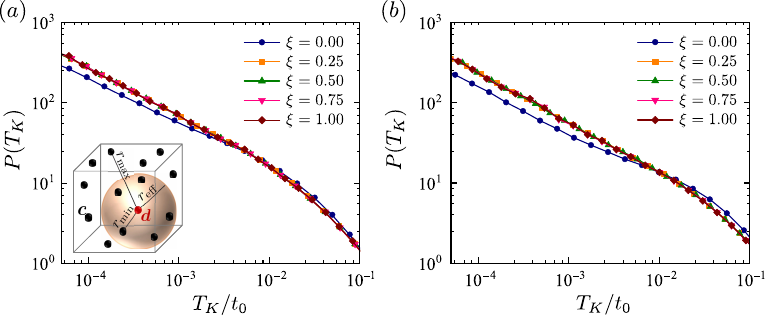}
\par\end{centering}
\caption{\label{fig:radius} Convergence of the distribution of single-impurity Kondo temperatures $P(T_K)$ as a function of the effective radius $r_{\mathrm{eff}}$ around the local moment, with $ r_{\mathrm{eff}} = r_{\mathrm{min}} + \xi (r_{\mathrm{max}} - r_{\mathrm{min}})$, where $ r_{\mathrm{min}} $ and $r_{\mathrm{max}} $ represent, respectively, the closest and farthest distances between the local moment and a conduction electron site. Here $L=30$, and (a) $n=0.010$ and (b) $n=0.020$. Inset: pictorial representation of $r_{\mathrm{eff}} $. }
\end{figure}

Importantly, for each single-impurity Kondo problem, we connect the corresponding local moment to all sites in the electronic bath.  Because the hopping $t_{ij}$ decay exponentially with the distance,  we expect that the contributions from conduction sites far away from the impurities are negligible, meaning that only a subset of conduction band sites (within an effective radius $r_{\mathrm{eff}}$) significantly interact with the local moment. To speed up our calculations, we implement the following scheme. Let $r_{\mathrm{min}}$ and $r_{\mathrm{max}}$ represent, respectively, the closest and farthest distances between the local moment and a conduction electron site.  We define the effective radius as $ r_{\mathrm{eff}} = r_{\mathrm{min}} + \xi (r_{\mathrm{max}} - r_{\mathrm{min}})$. By this definition, $\xi=0$ corresponds to a local coupling, whereas $\xi=1$ corresponds to the local moments interacting with all conduction sites. We find that $ \xi=0.25$ already captures the relevant low-energy physics [see Fig.~\ref{fig:radius}(a-b)], and we therefore set $\xi=0.25$ in all the results presented in this work.

\section{Local moment susceptibility}

In the limit $n\to0$,  where we decouple the local moments from the conduction electrons,  it is possible to obtain an analytical expression for the local moment susceptibility.  In the original Bhatt-Lee solution, all local moments within the system interact with one another. As one removes the strongly coupled singlet, all remaining couplings are renormalized \citep{bhatt_lee82}. These renormalized couplings are generically smaller in magnitude with respect to the original ones, implying they affect the properties of the systems at much lower energy scales, where $Q\left(J_{ab}\right)$ is already singular.  Moreover, the renormalization of the remaining couplings is an effect of order $1/\mathcal{N}$ for SU$\left(\mathcal{N}\right)$ spins \citep{hoyos04}, making the geometrical decimation consistent with our large$-\mathcal{N}$ solution of the problem. 

Using the geometrical procedure, we numerically calculate the spin susceptibility and write $\chi\left(T\right)=\nf/T$. We observe that $\chi\left(T\right)$ inherits the singularity of $Q\left(J_{ab}\right)$ at low$-T$,  with the singularity being more pronounced for small impurity density. To model the numerical data, one assumes that each spin defines a sphere of volume $v\propto\ell^{3}$, where $\ell \sim n^{-1/3}$ is the typical inter-spin distance. Since all these spheres touch each other, they are on the verge of forming a singlet but are still free. Therefore, if one relates the length scale $\ell$ with energy via $J_{ab} \sim \exp{\left(-2r_{ab}/a\right)}$, we can write $\chi\left(T\right)\sim1/T\ln^{3}\left(t_{0}/T\right)$ \citep{zhou09}, which is a logarithmic correction to the free moment result. Although this simple formula accurately describes the numerical data \citep{zhou09}, we refrain from implementing it here, as the effective description of the susceptibility singularity as a power law captures the data well over the full range of $n$ we study. 

We use two complementary methods to extract the power-law exponent $\alpha$. First, we extract the exponent fitting the slope of the linear behavior of $P\left(T_{K}\right)$ on a log-log scale. Complementarily,  we extract $\alpha$ directly from the data \citep{clauset09}. We assume that $P\left(T_{K}\right)$ displays a power-law form in the interval $T_{K}<T_{K}^{{\rm max}}$, with $T_{K}^{{\rm max}}$ the upper cutoff scale below which the power-law holds. The value of $T_{K}^{{\rm max}}$ for the power-law model that is most likely to have generated this data is given by $\alpha=1-\left\langle \ln\left(T_{K}^{{\rm max}}/T_{K}\right)\right\rangle _{T_{K}<T_{K}^{{\rm max}}}^{-1}$ \citep{clauset09}. In practice, we plot $\alpha\times T_{K}^{{\rm max}}$ and select the optimal $T_{K}^{{\rm max}}$ inside an energy window where $\alpha$ is reasonably stable. Both methods agree, enhancing confidence in the value of the exponent.

%%%%%%%%%%%
